\documentclass{jfm}
\usepackage{graphicx}
\usepackage{epstopdf, epsfig}
\usepackage{amsmath}

\newcommand{\myparI}[2]{\frac{\partial #1}{\partial #2}}
\newcommand{\myparII}[2]{\frac{\partial^2 #1}{\partial #2^2}}

\newcommand{\mydiff}[2]{\frac{\mathrm{d} #1}{\mathrm{d} #2}}
\newcommand{\mydiffII}[2]{\frac{\mathrm{d}^2 #1}{\mathrm{d} #2^2}}

\newcommand{\mymatdiff}[2]{\frac{\mathrm{D} #1}{\mathrm{D} #2}}

\shorttitle{Free-stream coherent structures in parallel compressible boundary-layer flows}
\shortauthor{E. C. Johnstone and P. Hall}

\title{Free-stream coherent structures in parallel compressible boundary-layer flows at subsonic Mach numbers}

\author{Eleanor C. Johnstone\aff{1}
	\corresp{\email{eleanor.johnstone@manchester.ac.uk}}
	\and Philip Hall\aff{2}}

\affiliation{\aff{1}Department of Mathematics, University of Manchester, Manchester M13 9PL, UK
	\aff{2}School of Mathematics, Monash University, Melbourne, VIC 3800, Australia}

\begin{document}

\maketitle

\begin{abstract}
As a first step towards the asymptotic description of coherent structures in compressible shear flows, we present a description of nonlinear equilibrium solutions of the Navier--Stokes equations in the compressible asymptotic suction boundary layer (ASBL). The free-stream Mach number is assumed to be $< 0.8$ so that the flow is in the subsonic regime and we assume a perfect gas. We extend the large-Reynolds number free-stream coherent structure theory of \cite{deguchi_hall_2014a} for incompressible ASBL flow to describe a nonlinear interaction in a thin layer situated just below the free-stream which produces streaky disturbances to both the velocity and temperature fields, which can grow exponentially towards the wall. We complete the description of the growth of the velocity and thermal streaks throughout the flow by solving the compressible boundary-region equations numerically. We show that the velocity and thermal streaks obtain their maximum amplitude in the unperturbed boundary layer. Increasing the free-stream Mach number enhances the thermal streaks, whereas varying the Prandtl number changes the location of the maximum amplitude of the thermal streak relative to the velocity streak. Such nonlinear equilibrium states have been implicated in shear transition in incompressible flows; therefore, our results indicate that a similar mechanism may also be present in compressible flows.
\end{abstract}

\begin{keywords}
\end{keywords}

\section{Introduction}\label{sec:intro}
It is known since \citet*{kline_reynolds_schraub_runstadler_1967} that transitionally turbulent flows exhibit clear structure within the boundary layer in the form of streaky flows coupled to roll flows in the plane perpendicular to the unperturbed flow. Recent understanding of these structures has been aided by the identification of three-dimensional, nonlinear equilibrium solutions of the Navier--Stokes equations in canonical shear flows. These states, now commonly known as exact coherent structures, have underlying physics described by the vortex-wave interaction mechanism \citep{hall_smith_1991, hall_sherwin_2010} in which forcing in the critical layer of the wave drives a roll flow which produces a streak; the streaky flow is then itself unstable to the wave. This tripartite interaction is also known as a self-sustaining process \citep{waleffe_1997}. Exact coherent structures have been found in a wide range of shear flows both computationally (see, for example, \citealt*{faisst__eckhardt_2003}; \citealt*{waleffe_2001, waleffe_2003}; \citealt*{wedin_kerswell_2004}; \citealt*{wang_gibson_waleffe_2007}) and asymptotically (\citealt*{hall_sherwin_2010}; \citealt*{deguchi_hall_2014a, deguchi_hall_2014b}).

More recently, a series of papers by Deguchi and Hall (\citealt*{deguchi_hall_2014a, deguchi_hall_2014b, deguchi_hall_2015a, deguchi_hall_2015b, deguchi_hall_2018}) develop a quite distinct asymptotic theory to describe coherent states where the nonlinear wave-roll-streak interaction takes place in a layer at the edge of the free-stream; these states are termed free-stream coherent structures. Their existence relies on the exponential approach of the boundary-layer flow to its free-steam form. The layer where the interaction takes place is termed the production layer, and the behaviour of the flow in the region between the production layer and the wall (termed the adjustment layer) is found to vary depending on the flow configuration: the behaviour for parallel boundary-layer flows (\citealt*{deguchi_hall_2014a, deguchi_hall_2014b}) is relatively simple compared to the complex asymptotic structure observed in non-parallel flows (\citealt*{deguchi_hall_2015a, deguchi_hall_2018}). In all configurations, the exponential form of the basic flow allows the streak disturbance, which is produced by a tiny nonlinear interaction in the free-stream involving waves, rolls and streaks, to grow exponentially beneath the production layer before obtaining its maximum amplitude in the adjustment layer. 

Thus it is hypothesised that free-stream coherent structures may play a key role in linking the inner and outer coherent structures seen in transitionally turbulent fluids \citep{deguchi_hall_2014a}. This detail would be particularly relevant in the context of jet acoustics for compressible flows, when disturbances originating in the free-stream may be implicated in the high frequency `screeching' observed in high-speed jet flows \citep{deguchi_hall_2018}.\\

 The study of coherent structures described above is focussed entirely on incompressible flows. However, the behaviour of transitionally turbulent compressible fluids is of vital importance to many industrial flows, particularly in the fields of aerospace engineering and acoustics. Past experimental and numerical studies have focussed on laminar-turbulent transition in compressible boundary layers in the context of the effect of free-stream vortical disturbances, with particular focus on bypass transition (see, for example, \citealt*{laufer1954factors}; \citealt*{kendall1975wind}; \citealt*{demetriades1989growth}; \citealt*{graziosi2002experiments}; \citealt*{mayer2011direct}). By extending the incompressible theory of \citet*{leib1999effect}, \cite{ricco2007response} show that free-stream vortical disturbances can induce temperature fluctuations that lead to the formation of `thermal streaks'; the growth of these streaks is enhanced at larger free-stream Mach numbers \citep*{marensi2017nonlinear}. 
 
 There has been little work, however, into the asymptotic description of coherent structures in the context of compressible flows; previous asymptotic studies of coherent structures have been confined to incompressible, canonical shear flows in channels and pipes where the only key parameter governing the dynamics is the Reynolds number. The study of exact coherent structures in two-parameter space has previously only been conducted in the context of stably-stratified flows (\citealt*{eaves2015disruption}; \citealt*{deguchi2017scaling}; \citealt*{lucas2017irreversible}; \citealt*{lucas2017layer}; \citealt*{olvera2017exact}), where it is shown that the Prandtl number plays a key role in the structure of the states found \citep*{langham2020stably}. 
 
 However, the organised streaky structures observed experimentally in incompressible flows have been identified in supersonic compressible flows both experimentally (for a thorough review see \citealt*{spina1994physics}) and numerically (\citealt*{ringuette2008coherent}; \citealt*{pirozzoli2008characterization}). The structures found are consistent with the hairpin loop model of wall turbulence, with low-speed, elongated streaks observed in the logarithmic region. Thus there exists compelling evidence for the similarity between compressible and incompressible coherent structures. Indeed, the main effect of compressibility in turbulent shear boundary layers lies in the density fluctuations \citep{morkovin1962effects}, and it is generally accepted that for moderate free-stream Mach numbers $M_\infty <~ 2$, the dynamics of compressible shear boundary layers do not differ greatly from their incompressible counterparts \citep{spina1994physics}. \\

We now investigate free-stream coherent structures in parallel asymptotic suction boundary-layer (ASBL) flow at subsonic Mach numbers ($M_\infty < 0.8$), under the perfect gas assumption. The key questions to consider are: (a) can we confirm that the asymptotic free-stream coherent structures found in compressible shear flows are related to those found in their incompressible counterparts?; and (b) what is the influence of the additional physical parameters, namely the Mach number $M_\infty$ and the Prandtl number $\sigma$? Of particular importance is the determination of the relative influence of the Mach number and the Reynolds number. 

For the compressible problem, we find that the location of the production layer now depends on both the Prandtl number and the free-stream Mach number, in addition to the Reynolds number. As in the incompressible problem, the nonlinear interaction in the production layer produces a disturbance to the streamwise velocity field (a `streak') that grows exponentially down towards the wall through interaction with the mean flow. However, the nonlinear interaction also induces a disturbance to the temperature field, a `thermal streak', which also grows exponentially down towards the wall. The amplitude of the thermal streaks is enhanced as the Mach number is increased. At the wall, both the velocity and thermal streaks vanish to satisfy the wall boundary conditions. The location where the thermal streak attains its maximum amplitude relative to the velocity streak is controlled by the Prandtl number; at higher Prandtl numbers (where momentum diffusivity dominates thermal diffusivity) the thermal streak attains its maximum closer to the wall than the velocity streak.\\

The compressible production layer problem for ASBL flow is asymptotically more complex than the incompressible problem due to the additional equations and parameters. However, a key feature is that the production-layer equations for the velocity field are the same as for the incompressible problem. The heat equation is then driven by this velocity field. This effect arises due to the location of the thin production layer being just below the free-stream, where compressibility effects are negligible. Importantly, this result also means that the solution of the nonlinear eigenvalue production-layer problem, which was computed by direct numerical simulation by \cite{deguchi_hall_2014a}, can also be used for the compressible problem. This significantly reduces the amount of computational work required. However, as discussed at the end of this paper, we expect that this reduction will not hold in general for other compressible regimes at higher free-stream Mach number due to the presence of non-parallel effects such as shocks. \\

The rest of this paper is presented as follows: in \S\ref{sec:fcsinc}, we provide a brief description of free-stream coherent structures in incompressible ASBL flow. We then define the governing equations for compressible ASBL flow in \S\ref{sec:govern} and find the basic flow in \S\ref{sec:basicflow}. The production layer problem is then described in \S\ref{sec:production}. We present the solution below the production layer and down to the wall in \S\ref{sec:adjustment}. We then present results for a variety of parameters in \S\ref{sec:results} and finally in \S\ref{sec:discussion} we draw some conclusions.

\section{Free-stream coherent structures in incompressible parallel boundary-layer flows}\label{sec:fcsinc}
To provide some context for the discussion of free-stream coherent structures in the compressible asymptotic suction boundary layer (ASBL) flow, we briefly summarise the results of \cite{deguchi_hall_2014a} for free-stream coherent structures in incompressible ASBL flow.\\

Incompressible ASBL flow describes viscous, incompressible flow, $(u^*, v^*, w^*)$ with respect to Cartesian coordinates $(x^*, y^*, z^*) = 0$, with viscosity $\mu$, over a flat plate at $y^* = 0$. Uniform flow exists in the free-stream, so denoting free-stream values by subscript $\infty$, at the free-stream $(u^*, v^*, w^*) = (u_\infty, -v_\infty, 0)$. The plate is subject to constant suction, so the velocity at the plate is $(u^*, v^*, w^*) = (0, -v_\infty, 0)$. Non-dimensionalising the velocity components on the free-stream speed $u_\infty$ and the coordinates on the length scale $\mu/v_\infty$, and defining the Reynolds number $\Rey = u_\infty/v_\infty$, the basic flow is given by
\begin{equation}
(u_b, v_b, w_b) = ( 1-\rm e^{-y}, -\Rey^{-1}, 0).
\end{equation}
\cite{deguchi_hall_2014a} showed that at high Reynolds numbers, the incompressible Navier--Stokes equations allow for nonlinear equilibrium solutions taking the form of a roll-wave-streak interaction propagating in a viscous layer at the edge of the boundary layer; this layer is termed the production layer and the solutions are known as free-stream coherent structures. The interaction in the production layer is characterised by nonlinear travelling wave solutions propagating with almost free-stream speed, with streamwise and spanwise length scales comparable to those of the boundary-layer. Seeking a solution with these scalings shows that the production layer in ASBL flow is located at $y = \ln \Rey$.

The solution inside the production layer $\boldsymbol U(X, Y, z) = (U, V, W)$, where $(x, y, z) = (X-ct, y-\ln \Rey, z)$, $c =1-\Rey^{-1} c_1$, is determined by numerically solving the full Navier--Stokes equations at unit Reynolds number as a nonlinear eigenvalue problem for the wave speed $c_1$ of the travelling wave:
\begin{eqnarray}
&([\boldsymbol U+c_1\boldsymbol{\hat i}]\bcdot \bnabla)\boldsymbol U = -\bnabla P + \nabla^2\boldsymbol U, \label{eq:incPLmom} \\
& \bnabla \bcdot \boldsymbol U= 0. \label{eq:incCont}
\end{eqnarray}
The asymptotic structure of the solution emerging from the lower side of the production layer shows that below the layer the disturbance to the streamwise velocity (termed the streak), which occurs as a result of the nonlinear interaction in the production layer, can grow exponentially while the other velocity components decay. Thus moving beneath the production layer,
\begin{eqnarray}
&\displaystyle u \to 1 -  \mathrm e^{-y} + \frac{d_0}{\Rey} + \frac{ J_1}{\Rey^{\omega_1}}\mathrm e^{(\omega_1-1)y}\cos(2\beta z) + \frac{ J_1 K_1}{\Rey^{2\omega_1}4\omega_1}e^{(2\omega_1-1)y}+ \cdots, \label{eq:incvelsol1} \\
&\displaystyle v \to -\frac 1 \Rey + \frac{ K_1}{\Rey^{\omega_1}} \mathrm e^{\omega_1y} \cos(2\beta z) + \cdots, \label{eq:incvelsol2}
\end{eqnarray}
where
\begingroup
\def\thesubequation{\theequation \textit{a--b}}
\begin{subeqnarray}
	J_1 = \frac{K_1}{(\omega_1-1)^2 + (\omega_1-1) - 4 \beta^2}, \quad \omega_1 = \frac{ -1 + \sqrt{ 1 + 16\beta^2}}{2} \geq 0, 
\end{subeqnarray}
\endgroup
for spanwise wavenumber $\beta$, and where $K_1$ is found as part of the numerical solution of the eigenvalue problem in the production layer. By solving for the induced flow throughout the boundary-region between the production layer and the wall, \cite{deguchi_hall_2014a} show that for $\beta < 1/\sqrt{2}$ the streak disturbance grows down to the main part of the boundary layer, before being reduced to zero at the wall to satisfy the near-wall boundary conditions. 

\section{Governing equations for compressible asymptotic suction boundary layer (ASBL) flow}\label{sec:govern}
 We now consider the compressible counterpart of ASBL flow. Consider a viscous, compressible perfect gas with dimensional density, temperature and viscosity $\rho^*$, $\theta^*$ and $\mu^*$ respectively, flowing with velocity $\boldsymbol{u^*} = (u^*, v^*, w^*)$ with respect to Cartesian co-ordinates $(x^*,   y^*, z^*)$ over an infinitely long flat plate at $y^* =0$. Uniform suction exists at the plate boundary so that, denoting free-stream values by subscript $\infty$, the velocity is $\boldsymbol{u^*} = (0, -v_\infty, 0)$ at the plate and $\boldsymbol{ u^*} \to (u_\infty, -v_\infty, 0)$ a long way from the plate. The suction at the plate does not allow for zero heat transfer over the plate due to the transfer of kinetic energy across the plate, and therefore we assume the temperature at the plate is fixed so that $\theta^* = \theta_p$ at $y^* = 0$.

We non-dimensionalise by scaling the coordinates $(x^*,   y^*, z^*)$ on the velocity-boundary-layer thickness $\delta = \mu_\infty /\rho_\infty v_\infty$, the velocity components $(u^*, v^*, w^*)$ on $u_\infty$, the pressure on $\rho_\infty u_\infty^2$ and the quantities $\rho^*$, $ \theta^*$ and $\mu^*$  on their free-stream values. We define the Reynolds number $\Rey$ by
 \begin{equation}
 \Rey = u_\infty/v_\infty.
 \end{equation}
 Throughout the analysis that follows, we assume the Reynolds number is large. We also define the following physical constants:
\begin{itemize}
	\item $c_v$, $c_p,$ are the specific heat at constant volume and constant pressure respectively,
	\item $\gamma = c_p/c_v$ is the ratio of specific heats; for air $\gamma \approx 1.4$,
	\item $a_\infty^2 = (\gamma-1) c_p \theta_\infty$ is the square of the speed of sound in the free-stream,
	\item $M_\infty = u_\infty/a_\infty$ is the free-stream Mach number which we shall assume is in the subsonic regime throughout our analysis so that $M_\infty < 0.8$,
	\item $k$ is the thermal diffusivity of the gas,
	\item $\sigma = \mu_\infty c_p/k$ is the Prandtl number which defines the ratio of momentum diffusivity to thermal diffusivity; for air, $\sigma \approx 0.71$,
	\item $R$ is the molecular gas constant. 
\end{itemize}
We choose parameters $\gamma$ and $\sigma$ that are appropriate for the ideal gas assumption; in particular, this means that $\sigma < 2$, which will become important in the scaling arguments below.

Then, using mixed notation so that $(x_1,x_2,x_3)$ represents $(x,y,z)$, $\bnabla=( \p_{x_1}, \p_{x_2}, \p_{x_3})$ and $\boldsymbol u = ( u_1, u_2, u_3)$ represents $( u,  v,  w)$, the Navier--Stokes equations have the form 
\begin{eqnarray}
&\displaystyle  \rho \mymatdiff{u_i}{t } = -\myparI{ p}{x_i} + \frac{1}{\Rey} \left\{ \myparI{}{ x_i} \left( -\frac{2}{3}\mu \bnabla \bcdot \boldsymbol{u} \right) + \myparI{}{x_j}\left(  \mu \myparI{ u_j}{ x_i} \right) + \myparI{}{ x_j}\left( \mu \myparI{u_i}{ x_j} \right)\right\} \ (i, j = 1, 2, 3), \nonumber\\
\label{eq:PLmomen}  \\
&\displaystyle \myparI{ \rho}{ t} +  \bnabla \bcdot ( \rho \boldsymbol{ u}) = 0, \label{eq:PLcont} \\
&\displaystyle \rho \mymatdiff{\theta}{t } = \frac{(\gamma -1)M_\infty^2}{\Rey} \Phi  + (\gamma -1)M_\infty^2 \mymatdiff{ p}{t } + \frac{1}{\Rey}\frac{1}{\sigma}\myparI{}{x_i} \left(\mu \myparI{ \theta}{ x_i} \right), \label{eq:PLheat} \\
&\displaystyle p =  \theta_\infty U_\infty^{-2} \rho R   \theta,  \label{eq:PLstate}   
\end{eqnarray}
where the dissipation function $\Phi$ is defined by
\begin{equation}\label{eq:Phind} 
\Phi = \frac{1}{2} \mu e_{ij}  e_{ij}  -\frac{2}{3} \mu ( {\bnabla} \bcdot \boldsymbol{ u})^2,
\end{equation}
and $e_{ij} = \p u_i/ \p x_j + \p u_j/ \p x_i$ is the rate of strain tensor.

We close the equations of motion using the Chapman--Rubesin constitutive law \citep{chapman1949temperature} which assumes a linear relationship between velocity and temperature; after non-dimensionalisation this reduces to \begin{equation}\label{eq:const}
\mu = \theta.
\end{equation}
 
\section{The basic flow}\label{sec:basicflow}
We now solve the equations of motion for the basic boundary-layer flow state. The asymptotic suction boundary layer (ASBL) flow is steady, two-dimensional and independent of $x$. Therefore we seek a boundary-layer solution in the form
\begin{subeqnarray}\label{eq:blexp}
&\left(u, v, w, p \right) = \left(\hat u(y),\ \Rey^{-1} \hat v(y), \ 0 , \ \hat p(y) \right),\\
&\left(\theta, \rho, \mu \right) = \left(\hat \theta(y), \  \hat \rho(y),\  \hat \mu(y)\right),
\end{subeqnarray}
where the scaling for the normal velocity arises from the need to retain viscous effects in the boundary layer. The boundary conditions at the plate and the free-stream are given by
\begin{subeqnarray}
	& (\hat u, \hat v, \hat w) = \left(0, -1, 0\right), \quad  \hat \theta = \theta_p/\theta_\infty \ \mbox{at} \  y = 0,\label{eq:boundaryconds} \\
	& (\hat u, \hat v, \hat w)  \to \left(1, -1 , 0\right), \quad \hat p \to p_\infty/ \rho_\infty u_\infty^2 , \quad (\hat \theta, \hat \rho, \hat \mu) \to (1, 1, 1) \ \mbox{as} \   y \to \infty.\label{eq:boundarycond0s}
\end{subeqnarray}

We substitute the expansion \eqref{eq:blexp} into the governing equations (\ref{eq:PLmomen})--(\ref{eq:PLstate}) and, assuming that the Reynolds number is large, retain leading order terms. The $y$-momentum equation from (\ref{eq:PLmomen}) with $i = 2$ reduces to $\partial \hat p /\partial y=0$, which means that the pressure $\hat p$ is constant across the boundary layer and equal to its free-stream value of $p_\infty/ \rho_\infty U_\infty^2$. It follows that the equation of state (\ref{eq:PLstate}) reduces to $\hat \rho \hat \theta = 1$, and by the Chapman--Rubesin law (\ref{eq:const}), we then obtain $\hat \rho \hat \mu = 1$. Then the continuity equation (\ref{eq:PLcont}) reduces to $\partial_y(\hat \rho \hat v) = 0$; integrating and applying free-stream boundary conditions (\ref{eq:boundarycond0s}) gives $\hat \rho \hat v = -1$ across the boundary layer. Thus, $\hat v = -\hat \theta$, so in particular, the suction condition at the plate gives $\theta_p/\theta_\infty =  1$.

We now use the Dorodnitsyn--Howarth transformation \citep{dorodnitsyn1942boundary, howarth1948concerning} given by
\begin{equation}\label{eq:dhtrans}
\xi = \int_0^y  \hat \rho(y') \ \mathrm{d}y',
\end{equation}
so that $y$-derivatives $\rm d_y$ are replaced by $\hat \rho(\xi) \rm d_\xi$. The equations of motion then reduce to
\begingroup
\def\thesubequation{\theequation \textit{a--b}}
\begin{subeqnarray}\label{eq:basicfloweqns}
-\hat u' = \hat u'', \quad -\hat \theta' = \sigma^{-1} \hat \theta'' + (\gamma-1)M_\infty^2 (\hat u')^2,
\end{subeqnarray} 
\endgroup
where prime denotes derivative with respect to $\xi$. Thus the full basic solution is given by
\begingroup
\def\thesubequation{\theequation \textit{a--b}}
\begin{eqnarray}
&\displaystyle \hat u(\xi) = 1-\mathrm e^{-\xi},  \quad
\hat \theta(\xi)= 1  +\frac{(\gamma-1)M_\infty^2 \sigma}{2(2-\sigma)} (\mathrm e^{-\sigma \xi} -\mathrm e^{-2\xi}), \label{eq:basicflowsol}\\
& \hat v (\xi)  = -\hat \theta(\xi), \quad \hat w(\xi) = 0, \quad \hat p(\xi) =p_\infty/ \rho_\infty u_\infty^2, \label{eq:basicflowsol4}\\
& \hat \rho(\xi)  = \left(\hat \theta(\xi)\right)^{-1}, \quad \hat \mu(\xi) = \hat \theta(\xi). \label{eq:basicflowsol3}
\end{eqnarray}
\endgroup
We can invert the Dorodnitsyn-Howarth transformation (\ref{eq:dhtrans}) as
\begin{equation}\label{eq:dhinversion}
y  = \int_0^\xi \hat\theta(\xi') \ \mathrm d \xi'=  \xi - \frac{(\gamma-1)M_\infty^2 \sigma}{2(2-\sigma)}\left( \frac{1}{\sigma}\mathrm e^{-\sigma \xi} - \frac{1}{2}\mathrm e^{-2\xi} -\frac{1}{\sigma}  + \frac{1}{2}\right).
\end{equation}
Thus if $\xi$ is small, then $y \approx \xi$ close to the wall, and if $\xi$ is large, then we can approximate the Dorodnitsyn-Howarth variable by
\begin{eqnarray}
&\displaystyle \xi \approx	g(y) = y + C_0; \quad \displaystyle C_0=  \frac{(\gamma-1)M_\infty^2 \sigma}{2(2-\sigma)} \left(-\frac{1}{\sigma}  + \frac{1}{2}\right). \label{eq:c0}
\end{eqnarray}
Consequently, both at the wall and the free-stream, we can write the basic flow in terms of the physical variable $y$. For the interior region we find $\xi = g(y)$ by solving the inversion equation (\ref{eq:dhinversion}) numerically.\\

Thus for large $\xi$, \textit{i.e.} large $y$, the basic streamwise velocity is given by $\hat u \approx 1-\mathrm e^{-C_0}\mathrm e^{-y}$. Thus the streamwise velocity approaches its free-stream form exponentially as a function of distance from the wall. Therefore the free-stream coherent structure theory of \cite{deguchi_hall_2014a} can be applied. The basic solution for the temperature field also approaches its free-stream form exponentially, with the rate of decay being dependent on the value of the Prandtl number. As discussed above in \S\ref{sec:govern}, gases which provide a good approximation to the ideal gas assumption have Prandtl numbers $\sigma < 2$, and therefore the decay of the basic state to its free-stream form will be dominated by the $\exp(-\sigma \xi)$ term in the basic flow (\ref{eq:basicflowsol}). Hence the decay will be slower than that of the streamwise velocity field $\hat u$ if $\sigma < 1$. Thus the thermal boundary layer is thicker than the velocity boundary layer if $\sigma < 1$, and vice-versa if $\sigma > 1$; this is consistent with laminar boundary layer theory which suggests that the thickness of the thermal boundary layer $\delta_\theta$ scales relative to the thickness of the velocity boundary layer $\delta_v$ as $\delta_\theta \sim \delta_v \sigma^{-1/3}$ \citep[p.~307]{schlichting1968boundary}.
\section{The production layer problem for compressible ASBL flow}\label{sec:production}
Using the inversion of the Dorodnitsyn--Howarth transformation for large $\xi$ (\ref{eq:dhinversion}), at the production layer we obtain $\xi \approx y + C_0$, and therefore the solution in the production layer can be expressed in terms of the physical variable $y$. To find the location of the production layer and the scalings of the flow components in the layer, following \cite{deguchi_hall_2014a}, we seek a travelling-wave solution propagating with almost the free-stream speed with wavelength comparable to the boundary-layer scalings of \S\ref{sec:basicflow} so that $\p_x \sim \p_y \sim \p_z \sim O(1)$. Then if viscosity is to play a role in the interaction, $v  \sim  O(\Rey^{-1})$, and then by the continuity equation (\ref{eq:PLcont}), $1-u \sim w \sim O(\Rey^{-1})$. To retain convection terms in the $x$-momentum equation (\ref{eq:PLmomen}) the $\rho(\p_t + u \p_x)$ term must also be $O(\Rey^{-1})$; this defines the wave dependence in the production layer. The pressure must then be $O(\Rey^{-2})$ to stay in play. 

The streamwise component of the velocity field in the production layer must include the basic flow component (\ref{eq:basicflowsol}) for matching. For large $\xi$, the basic flow $\hat u$ has the form $1-u = \exp(-y-C_0)$, therefore in the production layer, $\mathrm e^{-y-C_0} \sim \Rey^{-1}$. Thus the location of the production layer is given by $y = y_{PL} =  \ln \Rey-C_0$; this allows us to define a production-layer variable $Y = y-\ln \Rey+C_0$. The depth of the production layer must then be $O(1)$ to ensure that the streamwise velocity $u$ can only vary on an $O(1)$ length scale in the production layer. 

Thus a key feature of the compressible problem is that the location of the production layer depends on both the Prandtl number $\sigma$ and the free-stream Mach number $M_\infty$ through the constant $C_0$, in addition to the Reynolds number $\Rey$. Since $C_0 \propto M_\infty^2$, it is anticipated that the Mach number may have a strong influence on the hypersonic (large Mach number) production layer problem; this is discussed further in the conclusion. However, our choice of parameters means that $C_0 \ll \ln \Rey$. Therefore the values of $\sigma$ and $M_\infty$ do not strongly influence the location of the production layer. 

Under the scalings described above, the basic states for the streamwise velocity and temperature (\ref{eq:basicflowsol}) are given by
\begingroup
\def\thesubequation{\theequation \textit{a--b}}
\begin{subeqnarray}\label{eq:basicflowsol2}
	u = 1-\frac{1}{\Rey}\mathrm e^{-Y},  \quad
	\theta= 1  +\lambda\left( \frac{1}{\Rey^{\sigma}}\mathrm e^{-\sigma Y} - \frac{1}{\Rey^{2}}\mathrm e^{-2Y}\right),
\end{subeqnarray}
\endgroup
where 
\begin{equation}
\lambda = \frac{(\gamma-1)M_\infty^2 \sigma}{2(2-\sigma)}.
\end{equation}
Thus the largest deviation of the temperature field from its free-stream value at the production layer is controlled by the value of the Prandtl number $\sigma$. In particular, if $\sigma < 1$, then the deviation of the temperature field from its free-stream value is greater than the streamwise velocity deviation; this is again due to the relative thickness of the thermal and velocity boundary layers as discussed in \S\ref{sec:basicflow}.

It is also important to stress that although the $\exp(-\sigma \xi)$ exponential in the basic temperature state (\ref{eq:basicflowsol}) dominates the decay of the basic state to its free-stream value, upon exiting the production layer towards the wall as $Y \to -\infty$, any growing temperature disturbances will be dominated by the $\exp(-2Y)$ term in (\ref{eq:basicflowsol2}) and thus both exponentials need to be retained in the production-layer scalings.\\ 

Based on the discussion above, in the production layer we seek a solution of the Navier--Stokes equations in the form 
\begingroup
\def\thesubequation{\theequation \textit{a--e}}
\begin{subeqnarray}\label{eq:PLvars1}\displaystyle
	&(X, Y, z) = (x-ct, y-\ln \Rey+C_0, z); \quad
	c =  1 - \Rey^{-1}c_1 + \dots, \nonumber \\
	&\boldsymbol u  = (1, 0, 0) + \Rey^{-1}\boldsymbol{\bar u} (X, Y, z)+ \dots, \quad p = p_\infty/ \rho_\infty u_\infty^2 +  \Rey^{-2} \bar p(X, Y, z)  + \dots, \nonumber \\
	&(\theta, \rho, \mu) = 1 + \Rey^{-\sigma}(\bar \theta_1, \bar \rho_1, \bar \mu_1)(X, Y, z) + \Rey^{-2}(\bar \theta_2, \bar \rho_2, \bar \mu_2)(X, Y, z). 
\end{subeqnarray}
\endgroup
We substitute these scalings into the three-dimensional Navier--Stokes equations (\ref{eq:PLmomen})--(\ref{eq:PLstate}) and, at leading order, we obtain the production-layer problem:
\begin{eqnarray}
&\mathcal L\boldsymbol{\bar u} = -\bnabla \bar p + \nabla^2\boldsymbol{\bar u}, \label{eq:PLmom} \\
&\bnabla \bcdot \boldsymbol{\bar u}= 0, \label{eq:PLcontin} \\
&\mathcal L\bar \theta_1 = \sigma^{-1} \nabla^2 \bar \theta_1, \ \mbox{at order} \ \Rey^{-(\sigma +1)}, \label{eq:caseiHEAT1} \\
&\mathcal L \bar \theta_2 = (\gamma-1)M_\infty^2 \mathcal L \bar p  + (\gamma-1)M_\infty^2 \bar \Phi + \sigma^{-1} \nabla^2 \bar \theta_2, \ \mbox{at order} \ \Rey^{-3}, \label{eq:caseiHEAT2} \\
&\bar \rho_1 + \bar \theta_1 = 0, \ \mbox{at order} \ \Rey^{-\sigma}, \\ &\bar p = R\theta_\infty u_\infty^{-2} (\bar \rho_2 + \bar \theta_2), \ \mbox{at order} \ \Rey^{-2}, \label{eq:caseiSTATE}
\end{eqnarray}
where the operator $\mathcal L = ([\boldsymbol{\bar u}+c_1\boldsymbol{\hat i}]\bcdot \bnabla)$ and $\bnabla = (\p_X, \p_Y, \p_z)$.

We see that the production layer equations for the velocity field $\boldsymbol{\bar u}$ (\ref{eq:PLmom})--(\ref{eq:PLcontin}) are the same as the equations (\ref{eq:incPLmom})--(\ref{eq:incCont}) for the incompressible production layer problem in \cite{deguchi_hall_2014a}, which describe a unit-Reynolds-number eigenvalue problem for the wave speed $c_1$. Therefore the solution to this eigenvalue problem, which was calculated in \cite{deguchi_hall_2014a}, can now also be used for the compressible problem. The velocity field then drives the temperature field through the heat equations (\ref{eq:caseiHEAT1})--(\ref{eq:caseiHEAT2}); equation (\ref{eq:caseiHEAT1}), which is obtained at $O(Re^{-\sigma})$, is dominant in the production layer, but we require the solution of the equation at $O(Re^{-2})$ as the production layer is exited towards the wall. 

The production layer problem (\ref{eq:PLmom})--(\ref{eq:caseiSTATE}) is solved subject to boundary conditions 
\begin{eqnarray}
	&\boldsymbol{\bar u} \to (0, -1, 0), \quad \bar \theta_1 \to \lambda  \mathrm e^{-\sigma Y} , \quad \bar \theta_2 \to  -\lambda  \mathrm e^{-2 Y}   \quad \mbox{as} \ Y \to \infty, \label{eq:bc1}\\
&\boldsymbol{\bar u} \to (- \mathrm e^{-Y}, -1, 0), \quad
\bar \theta_1 \to \lambda   \mathrm e^{-\sigma Y},  \quad \bar \theta_2 \to -\lambda  \mathrm e^{-2 Y} \quad \mbox{as} \ Y \to -\infty, \label{eq:bc2}
\end{eqnarray}
and periodicity conditions; defining $\alpha$ and $\beta$ as the streamwise and spanwise wavenumbers respectively, 
\begin{eqnarray}
&( \boldsymbol{\bar u}, \bar \theta_{1,2})(X, Y, z)  = ( \boldsymbol{\bar u}, \bar \theta_{1,2})(X + 2\upi/\alpha, Y, z), \label{eq:caseiiBCspan}\\
&(\boldsymbol{\bar u}, \bar \theta_{1,2})(X, Y, z)  = ( \boldsymbol{\bar u}, \bar \theta_{1,2})(X , Y, z+ 2\upi/\beta). \label{eq:caseiBCspan}
\end{eqnarray}
Thus, considering the second of these boundary conditions (\ref{eq:caseiBCspan}), we see that as in the incompressible problem, the boundary conditions allow for the streamwise velocity disturbance to grow exponentially beneath the production layer. However, the conditions also allow for a disturbance to the temperature field to grow exponentially and at a faster rate than the streamwise velocity disturbance. Coming out of the production layer $\bar \theta_2$ is dominant, however the $\bar \theta_1$ part of the disturbance, which satisfies a homogeneous equation, must be retained as it is needed at the wall. All disturbances must be reduced to zero at the wall and therefore, as in the incompressible problem, the maximum value of the disturbances will occur in a layer between the wall and the production layer where the basic flow adjusts to accommodate the disturbance. 
\section{The adjustment layer problem}\label{sec:adjustment}
Below the production layer, the flow returns to the unperturbed boundary layer flow (\ref{eq:basicflowsol})--(\ref{eq:basicflowsol3}) at leading order. However, the nonlinear production layer interaction produces exponentially growing disturbances to the streamwise velocity and temperature fields that interact with the basic flow beneath the production layer. The flow between the production layer and the wall adjusts to accommodate the disturbances; we thus refer to this region as the adjustment layer. The solution in the upper part of this layer is dominated by the solution exiting the production layer. Then as the wall is approached, the solution is described by the full boundary-region equations. 
\subsection{The solution exiting the production layer}
Firstly, above the production layer as $Y \to \infty$, the velocity must eventually return to its free-stream form $\boldsymbol{\bar u} = (0, -1, 0)$. As in \cite{deguchi_hall_2014a}, the decay of the streamwise velocity $u$ will be proportional to $\rm e^{-Y-C_0}$, however the nonlinear interaction in the production layer gives a constant of proportionality which differs from unity. Thus the production layer interaction can give at most an $O(1)$ effect on the amplitude of the streamwise velocity displacement. Since the temperature field in the production layer is entirely driven by the equations for the velocity (\ref{eq:PLmom})--(\ref{eq:PLcontin}), any temperature disturbances will also decay above the production layer as there is no interaction to sustain them.\\

We now consider $Y \to -\infty$. To analyse the flow beneath the production layer, we decompose the velocity disturbance $\boldsymbol{\bar u}$ into vortex and wave components. The wave is associated with the $X$-dependent components of the velocity field. The $X$-independent components of the velocity are split into a roll flow, which is associated with the components $\bar v$ and $\bar w$, and the streak, which is the downstream velocity component $\bar u$. The combination of the roll and streak then makes a streamwise vortex. At leading order, the flow must satisfy the basic ASBL flow given by (\ref{eq:bc2}), and therefore we split the streak into a mean in $z$ and a $z$-dependent component (there is no mean in $z$ of the roll flow due to symmetry). We allow the $z$-independent term to grow exponentially in the adjustment layer as $Y \to -\infty$, but it must eventually reduce to the unperturbed basic flow. 

We decompose the temperature disturbances $\bar \theta_1$ and $\bar \theta_2$ in the same way; although it is not physically motivated, we have used the same vortex-wave notation for the temperature to allow better comparison with the velocity disturbance. Following the nomenclature outlined in \cite{ricco2007response}, we refer to the $X$ independent component of the temperature disturbance as a `thermal streak' and the corresponding streamwise velocity disturbance shall be termed a `velocity streak'. Hence, in the adjustment layer, we seek a solution in the form
\begin{eqnarray}
\boldsymbol{\bar u} &= (\overline{ u}_v(Y), -1, 0) + \boldsymbol u_v(Y, z) + \boldsymbol u_w(X, Y, z), \label{eq:PLexitVel}\\
&\bar \theta_{1,2} = \overline{ \theta}_{v_{1,2}}(Y) + \theta_{v_{1,2}}(Y, z) + \theta_{w_{1,2}}(X, Y, z), \label{eq:PLexitTemp1} 
\end{eqnarray}
where subscript $v$ refers to a vortex component and subscript $w$ refers to a wave component. 

As in the incompressible ASBL study of \cite{deguchi_hall_2014a}, outside of the production layer the roll flow decays as there is no longer any forcing from the Reynolds stresses in the critical layer of the wave to sustain it. The wave $\boldsymbol u_w$ also decays faster than the roll; this can be seen through a balance of advection--diffusion terms and is confirmed by the numerical results of \cite{deguchi_hall_2014a}. Since the temperature field is driven entirely by the velocity field, the same is true of the corresponding temperature components $\theta_{w_{1, 2}}$ and $\theta_{w_{1,2}}$. However, the velocity streak $\overline{u}_v + u_v$ can grow exponentially through interaction with the roll. The growth or decay of the velocity streak depends on the spanwise wavenumber $\beta$ through the periodicity conditions (\ref{eq:caseiiBCspan})--(\ref{eq:caseiBCspan}). The new feature for the compressible problem is that the interaction of the roll flow with the temperature field drives the growth of the thermal streak. \\

 We substitute the decomposition of the disturbances (\ref{eq:PLexitVel})--(\ref{eq:PLexitTemp1}) into the production-layer equations (\ref{eq:PLmom})--(\ref{eq:caseiSTATE}). After eliminating $\bar \rho$ and $\bar \mu$ by using (\ref{eq:caseiSTATE}) and the Chapman-Rubesin law (\ref{eq:const}), the resulting equations for the roll-velocity-streak flow are given by
\begin{eqnarray}
&\displaystyle  e^{-Y}v_v = \left(\myparII{}{Y} + \myparI{}{Y} + \myparII{}{z}\right)u_v \label{eq:PLexitXMOM} \\
&\displaystyle  \left(\frac{4}{3}\myparII{}{Y} + \myparI{}{Y} + \myparII{}{z}\right) v_v + \frac{1}{3}\frac{\partial^2w_v}{\partial z \partial Y} = 0, \label{eq:PLexitYMOM} \\
&\displaystyle  \left(\myparII{}{Y} + \myparI{}{Y} + \frac{4}{3}\myparII{}{z}\right) w_v + \frac{1}{3}\frac{\partial^2v_v}{\partial z \partial Y} = 0, \label{eq:PLexitZMOM} \\
&\displaystyle  \myparI{v_v}{Y} + \myparI{w_v}{z} = 0, \label{eq:PLexitCONT} \\
&\displaystyle \mydiff{\overline{u}_v}{Y} + \mydiffII{\overline{u}_v}{Y} = \frac{\beta}{2\pi} \mydiff{}{Y} \int_{z = 0}^{2\pi/\beta}  (u_v v_v) \ \mathrm d z, \label{eq:PLexitMEANVEL}
\end{eqnarray}
where the final equation for the mean velocity streak disturbance $\overline{u}_v$ has been found by taking the mean in $z$ of the production-layer $x$-momentum equation (\ref{eq:PLmom}). It is important to note the $\p_Y$ terms in the equations above which arise from the suction in the flow. It is these terms that allow the interaction of the mean part of the basic flow with the roll flow to produce growth.

The roll-velocity-streak equations (\ref{eq:PLexitXMOM})--(\ref{eq:PLexitMEANVEL}) are solved together with the equations for the thermal streak,
\begin{eqnarray}
&\displaystyle \myparI{\theta_{v_1}}{Y} + \frac{1}{\sigma}\myparII{\theta_{v_1}}{Y} + \frac{1}{\sigma}\myparII{\theta_{v_1}}{z} = -  v_v \sigma \lambda \mathrm e^{-\sigma Y},\label{eq:tempeqns1} \\
&\displaystyle \myparI{\theta_{v_2}}{Y} + \frac{1}{\sigma}\myparII{\theta_{v_2}}{Y} + \frac{1}{\sigma}\myparII{\theta_{v_2}}{z} = -2  \lambda v_v \mathrm e^{-2 Y} -2 (\gamma-1)M_\infty^2 \mathrm e^{-Y} \myparI{u_v}{Y}, \label{eq:tempeqns2} \\
&\displaystyle \mydiff{\overline{\theta}_{v_1}}{Y} + \frac{1}{\sigma}\mydiffII{\overline{\theta}_{v_1}}{Y} = \frac{\beta}{2\upi} \mydiff{}{Y} \int_{z = 0}^{2\upi/\beta}  (v_v \theta_{v_1}) \ \mathrm d z,\label{eq:tempeqns3} \\
&\displaystyle \mydiff{\overline{\theta}_{v_2}}{Y} + \frac{1}{\sigma}\mydiffII{\overline{\theta}_{v_2}}{Y} = \frac{\beta}{2\upi} \mydiff{}{Y} \int_{z = 0}^{2\upi/\beta} \left( v_v \theta_{v_2}  - (\gamma-1)M_\infty^2  \phi \right) \ \mathrm{d} z \label{eq:tempeqns4} ,
\end{eqnarray}
where
\begin{equation}
\phi = 4\left( \myparI{v_v}{Y} \right)^2 +  \left( \mydiff{\overline{u}_v}{Y} \right)^2  +\left( \myparI{u_v}{Y} \right)^2 + \left( \myparI{u_v}{z} \right)^2 + \left( \myparI{v_v}{z} \right)^2 + 2\myparI{v_v}{z} \myparI{w_v}{Y} + \left( \myparI{w_v}{Y} \right)^2.
\end{equation}\\

These equations are solved by Fourier expansion in $z$. Using the continuity equation (\ref{eq:PLexitCONT}) to define a stream function $\psi$ such that $v_v = \p_z \psi$ and $w_v = -\p_y \psi$, we combine the equations (\ref{eq:PLexitYMOM})--(\ref{eq:PLexitZMOM}) and seek a solution in terms of a Fourier expansion of $\psi$. The numerical results of \cite{deguchi_hall_2014a} show that the vortex wavelength is half that of the wave part of the flow, and therefore the wavelength of the vortex is $\pi/\beta$, which sets the wavenumbers of the Fourier expansion. Therefore, we seek a solution for $\psi$ in the form
\begin{subeqnarray}
&\psi = \sum_{n=0}^\infty a_n \cos(2n\beta z) + \sum_{n=1}^\infty b_n \sin(2n\beta z), \\
&a_n = \frac{2\pi}{\beta}\int_0^{\pi/\beta} \psi(Y,z) \cos(2n\beta z), \\
&b_n = \frac{2\pi}{\beta}\int_0^{\pi/\beta} \psi(Y,z) \sin(2n\beta z).
\end{subeqnarray}

The equations for $u_v$, $v_v$ and $\overline{u}_v$ are the same as those for the incompressible problem in \cite{deguchi_hall_2014a}, and thus we can use the solution obtained there given by (\ref{eq:incvelsol1})--(\ref{eq:incvelsol2}). Exiting the production layer, we still have the approximation $\xi = y + C_0$, and hence we can write the solution in terms of the physical variable $y$. Thus upon exiting the production layer, in original boundary layer coordinates  $(x, y, z)$ and associated flow quantities $(u, v, w)$, the solution is
\begin{eqnarray}
&\displaystyle u \to 1 -  \mathrm e^{-(y+C_0)} + \frac{d_0}{\Rey} + \frac{ J_1}{\Rey^{\omega_1}}\mathrm e^{(\omega_1-1)(y+C_0)}\cos(2\beta z) + \frac{ J_1 K_1}{\Rey^{2\omega_1}4\omega_1}e^{(2\omega_1-1)(y+C_0)}+ \cdots, \label{eq:velsol1} \nonumber \\
\\
&\displaystyle v \to -\frac 1 \Rey + \frac{ K_1}{\Rey^{\omega_1}} \mathrm e^{\omega_1(y+C_0)} \cos(2\beta z) + \cdots, \label{eq:velsol2}
\end{eqnarray}
where
\begingroup
\def\thesubequation{\theequation \textit{a--b}}
\begin{subeqnarray}
J_n = \frac{K_n}{(\omega_n-1)^2 + (\omega_n-1) - 4 n^2\beta^2}, \quad \omega_n = \frac{ -1 + \sqrt{ 1 + 16n^2\beta^2}}{2} \geq 0, 
\end{subeqnarray}
\endgroup
for $n \geq 1$. The constants $d_0$ and $K_1$ are found as part of the nonlinear eigenvalue production-layer problem, and were calculated for a range of $\beta$ in \cite{deguchi_hall_2014a}. The ``$\cdots$" represent more slowly growing harmonics in $z$. Thus as required, the flow returns to its unperturbed basic state at leading order, with exponentially-growing disturbances that can become as large as the basic flow in the main boundary region.\\

The solutions for $u_v$, $v_v$ and $\overline{u}_v$ are then used as forcing for the equations (\ref{eq:tempeqns1})--(\ref{eq:tempeqns4}) for the thermal streak. In original boundary-layer variables, so  $\theta  = 1 + \Rey^{-\sigma}\bar \theta_1 + \Rey^{-2}\bar \theta_2$, we find that upon exiting the production layer, 
\begin{eqnarray}\label{eq:tsol}
	&\displaystyle \theta \to 1 + \lambda \mathrm e^{-\sigma (y+C_0)} -  \lambda  \mathrm e^{-2(y+C_0)} +  \frac{d_1}{\Rey^{\sigma}} + \frac{d_2}{\Rey^2}  \nonumber \\
&\displaystyle \qquad  + \frac{1}{\Rey^{\omega_1}}\left( L_1 \mathrm e^{(\omega_1-\sigma)(y+C_0)} +  Q_1 \mathrm e^{(\omega_1 -2)(y+C_0)}\right)\cos(2\beta z) \nonumber \\
&\displaystyle \qquad\quad \quad + \frac{1}{\Rey^{2\omega_1}} \left( \frac{ L_1 K_1 \sigma}{4 \omega_1} \mathrm e^{(2\omega_1-\sigma)(y+C_0)} +  R_1 \mathrm e^{(2\omega_1-2)(y+C_0)}\right) + \cdots,
\end{eqnarray}
where
\begin{subeqnarray}
&\displaystyle L_n = \frac{-K_n \lambda\sigma}{(\omega_n - \sigma) + \sigma^{-1} (\omega_n-\sigma)^2 - \sigma^{-1}4n^2\beta^2},\\
&\displaystyle Q_n = \frac{-2 \lambda K_n - 2(\gamma-1)M_\infty^2 J_n (\omega_n-1) }{(\omega_n - 2) + \sigma^{-1} (\omega_n-2)^2 - \sigma^{-1}4n^2\beta^2}, \\
& \displaystyle R_n = -\frac 1 2 \,{\frac {{J_n}\, \left( {J_n}\,{{\omega_n}}^{3}-2\,{J_n}\,{{\omega_n}}^{2}-1/2\,{K_n}\, \right)  \left( \gamma-1\right) {M_\infty}^{2}\sigma}{ \left( \sigma+2\,{\omega_n}-2 \right) {\omega_n}}} \nonumber \\
&\displaystyle  -2\,{\frac {\sigma\, \left(  \left( {\beta}^{2}{n}^{2}+1/4 \right)\left( \gamma-1 \right) {M_\infty}^{2}{{J_n}}^{2}+1/4\,{M_\infty}^{2}{J_n}\,{K_n}\,\, \left( \gamma-1 \right) -1/4\,{K_n}\,{Q_n}\right) }{\sigma+2\,{\omega_n}-2}}
\end{subeqnarray}
The constants $d_1$ and $d_2$ are constants of integration and can be found by ensuring that the temperature solution satisfies the heat equation; $d_0$ is extracted from the numerical eigenvalue problem (\ref{eq:PLmom})--(\ref{eq:PLcontin}) and then $d_1$ and $d_2$ are found  using the velocity solution. \\

The asymptotic solution (\ref{eq:velsol1})--(\ref{eq:velsol2}) for $u$ and $v$ beneath the production layer shows that the roll flow always decays as the wall layer is approached, whereas the mean part of the velocity streak flow (\ref{eq:velsol1}) can grow beneath the production layer if $2\omega_1 < 1$, corresponding to values of $\beta < \sqrt 3/4$. The $z$-dependent part of the velocity streak can grow if $\omega_1 < 1$, corresponding to values of $\beta < 1/\sqrt 2$, and therefore these latter modes are the fastest growing. Thus if $\beta > 1/\sqrt 2$, then the velocity streak disturbance decays exponentially, and the nonlinear interaction in the production layer simply produces an $O(Re^{-1})$ correction to the flow.

Meanwhile, the asymptotic solution for the temperature (\ref{eq:tsol}) beneath the production layer shows that the thermal streaks can grow if $\omega_1  < \sigma $ or if $\omega_1 < 2$. For the range of values of Prandtl number $\sigma<2 $ considered, the modes proportional to $\exp( (\omega_1-2)(y+C_0))$ will dominate the growth, and therefore the nonlinear interaction in the production layer will always produce growing temperature disturbances for $\beta < \sqrt{3}/\sqrt{2}$. 

The structure of the solution with varying $\beta$ is summarised in table \ref{tab:betasig}. The asymptotic results suggest that there exists a case where the thermal streaks can grow independently while the velocity streak decays. However, solutions of the production layer problem (\ref{eq:PLmom})--(\ref{eq:PLcontin}) have not been found for values of $\beta \gtrsim 0.47$ \citep{deguchi_hall_2014a, deguchi_hall_2015a}, and therefore cases $(c)$ and $(d)$ are possibly not relevant.

 \begin{table}[t]
 	\centering	
 	\begin{tabular}{ c||c| c |c |c }
 		& $(a)$ $ \beta < \frac{\sqrt 3}{4}$ &$(b)$ $ \frac{\sqrt 3}{4}< \beta < \frac{1}{\sqrt 2}$ &$(c)$ $ \frac{1}{\sqrt 2} < \beta < \sqrt{\frac{3}{2}}$ &$(d)$ $ \beta > \sqrt{\frac{3}{2}}$ \\ 
 		\hline
 		$u_v$ & G& G &D & D\\ 
 		$\overline{u_v}$ & G & D &D &D\\ 
 		$\theta_v$ & G & G &G&D\\
 		$\overline{\theta_v}$ & G & G &D &D \\
	\end{tabular}
\caption{The growth and decay of the disturbances for different values of the spanwise wavenumber $\beta$. Growth is represented by `G' and decay by `D'. The growth and decay is shown for the both the mean in $z$ and the $z$-dependent parts of the flow.}
	\label{tab:betasig}
\end{table}
We see that in all cases, a nonlinear interaction in the production layer of size $O(\Rey^{-1})$, which drives $O(\Rey^{-2}), O(\Rey^{-\sigma})$ temperature perturbations, can induce much larger changes to the velocity and temperature fields of $O(\Rey^{\omega_1})$ in the main part of the boundary layer. We now consider the solution as it approaches the wall layer, where all disturbances are eventually reduced to zero to satisfy the wall conditions. 
\subsection{Boundary layer analysis}
The solutions exiting the production layer, (\ref{eq:velsol1}), (\ref{eq:velsol2}) and (\ref{eq:tsol}), do not satisfy the wall conditions. We now find the solution for the induced flow which is valid all the way down to the  wall. This solution should also match onto the solution exiting the production layer given by (\ref{eq:velsol1}), (\ref{eq:velsol2}) and (\ref{eq:tsol}). An examination of this solution shows that in the boundary-layer, disturbances can grow exponentially, becoming as large as the exponential part of the basic flow. The $z$-dependent part of the disturbance grows faster than the $z$-independent part; therefore, to match onto the solution exiting the production layer, the boundary-region solution will have $z$-dependence in the form of $\cos(2\beta z)$. However, the solution must also satisfy the wall boundary conditions (\ref{eq:boundaryconds}$a$), and therefore any disturbances must ultimately be reduced to zero at the wall. 

The solution in the boundary region is described in terms of the Dorodnitsyn-Howarth variable $\xi$; since disturbances are always small compared to the basic flow, the definition of the variable in equation (\ref{eq:dhtrans}) is valid throughout the flow, and in particular, $\partial_y = \hat \rho(\xi) \partial_\xi$. The inversion of the Dorodnitsyn-Howarth transformation for $y(\xi)$ is given in equation (\ref{eq:dhinversion}), however, unlike at wall and in the production layer, we cannot generally find an explicit relationship for $\xi(y)$ as it cannot be assumed that the exponential terms involving $\xi$ are smaller than the linear terms. Therefore, to find the solution for the physical variable $y$, we first solve the boundary-region equations in terms of $\xi$, and then use the monotonic relationship $y(\xi)$ in (\ref{eq:dhinversion}) to plot the solutions for each corresponding value of $y$. \\

Based on this discussion, in the boundary region we seek a solution in the form
\begin{subeqnarray}
	& u = \hat u(\xi) + \tilde u (\xi)\cos(2\beta z), \quad \hat v(\xi) = \Rey^{-1} \hat v(\xi) + \Rey^{-1} \tilde v(\xi)\cos(2\beta z), \\
	& w = \Rey^{-1} \tilde w(\xi)\sin(2\beta z), \quad p = \hat p(\xi) + \Rey^{-2} \tilde p(\xi)\cos(2\beta z), \\
	&\left(\theta, \rho, \mu \right) = \left( \hat \theta(\xi),  \hat \rho(\xi), \hat \mu(\xi) \right) + \left( \tilde \theta (\xi),  \tilde \rho(\xi), \tilde \mu(\xi) \right)\cos(2\beta z),
\end{subeqnarray} 
where the basic solution (hat quantities) is given by (\ref{eq:basicflowsol2}). We use the same velocity streak and thermal streak terminology to refer to the disturbances to the streamwise velocity and temperature fields respectively, and again the roll flow is associated with the disturbances to the $(v, w)$ components of the velocity. 

We substitute this expansion into the Navier--Stokes equations (\ref{eq:PLmomen})--(\ref{eq:PLstate}), which leads to a set of ordinary differential equations in $\xi$ for the leading-order disturbance amplitudes (tilde quantities). Following \cite{hall_1983}, we eliminate the pressure $\tilde p$ and the spanwise disturbance velocity $\tilde w$; then we also eliminate the viscosity $\tilde \mu$ and the density $\tilde \rho$ using the equation of state (\ref{eq:PLstate}) and the Chapman-Rubesin law (\ref{eq:const}). We are then left with three coupled differential equations for $\tilde u$ ( from the $x$-momentum equation), $\tilde v$ (from the $y$-momentum equation) and $\tilde \theta$ (from the temperature equation):
\begin{eqnarray}
	& A_1 \tilde u + A_2 \tilde u' + A_3 \tilde u'' = A_4 \tilde v + A_5 \tilde \theta + A_6 \tilde \theta',\label{eq:breXmom}\\
	& B_1 \tilde v + B_2 \tilde v' + B_3 \tilde v'' + B_4 \tilde v^{(3)} + B_5 \tilde v^{(4)} = B_6 \tilde \theta + B_7 \tilde \theta' + B_8 \tilde \theta'' + B_9 \tilde \theta^{(3)} + B_{10} \tilde \theta^{(4)} ,\label{eq:breYmom} \\
	& C_1 \tilde \theta + C_2 \tilde \theta' + C_3 \tilde \theta'' = C_4 \tilde u' + C_5 \tilde v.\label{eq:breTemp}
\end{eqnarray}
Here, superscript prime/number represents derivative in the usual way. The coefficients $A_k$, $B_k$ and $C_k$ depend on the basic solution and are too long to write here; details are available from the authors on request. These coupled equations are solved subject to zero-disturbance and no-slip boundary conditions at the wall, and matching to the solution exiting the production layer (\ref{eq:velsol1}), (\ref{eq:velsol2}), (\ref{eq:tsol}) at $\xi = \xi_{PL} = \ln \Rey$, so that
\begin{subeqnarray}
	& \tilde u(0) =0, \quad \tilde u(\xi_{PL}) = J_1 \Rey^{-\omega_1} \mathrm e^{(\omega_1-1)\xi_{PL}},\\
	&  \tilde v(0)  =\tilde v'(0)  = 0, \ \tilde v(\xi_{PL}) = K_1 \Rey^{-\omega_1} \mathrm e^{\omega_1\xi_{PL}}, \ \tilde v'(\xi_{PL}) = K_1 \omega_1 \Rey^{-\omega_1} \mathrm e^{\omega_1\xi_{PL}}, \\
	&  \tilde \theta(0) = 0, \quad \tilde \theta(\xi_{PL}) =  \Rey^{-\omega_1} \left( L_1 \mathrm e^{(\omega_1-\sigma)\xi_{PL}} + Q_1 \mathrm e^{(\omega_1-2)\xi_{PL}} \right).
\end{subeqnarray}
 The reduced boundary-region equations are now discretised on a grid with $N$ interior points and we use second-order accurate centred finite differences to approximate the derivatives. We denote the values of $\tilde u$, $\tilde v$ and $\tilde \theta$ at $\xi_i = (i-1)\Delta \xi$, $\Delta \xi = 1/N$ by $\tilde u(\xi_i) = \tilde u_i$, $\tilde v(\xi_i) = \tilde v_i$ and $\tilde \theta(\xi_i) = \tilde \theta_i$ respectively, where $0 \leq i \leq N+1$. The wall is at $\xi_1 = 0$ and the production layer is at $\xi_N = (N-1) \Delta \xi = \xi_{PL}$. 
 
 The discretised boundary-region equations are
 \begin{eqnarray}
 	& \alpha_1 \tilde u_{i+1} + \alpha_2 \tilde u_{i} + \alpha_3 \tilde u_{i-1} = \alpha_4 \tilde v_i + \alpha_5 \tilde \theta_{i+1} + \alpha_6 \tilde \theta_i + \alpha_7 \tilde \theta_{i-1}, \label{eq:xdisc}\\
 	& \beta_1 \tilde v_{i+2} + \beta_2 \tilde v_{i+1} + \beta_3 \tilde v_i + \beta_4 \tilde v_{i-1} + \beta_5 \tilde v_{i-2} = \beta_6 \tilde \theta_{i+2} + \beta_7 \tilde \theta_{i+1} + \beta_8 \tilde \theta_i + \beta_9 \tilde \theta_{i-1} + \beta_{10} \tilde \theta_{i-2} , \label{eq:ydisc}\\
 	& \gamma_1 \tilde \theta_{i+1} + \gamma_2 \tilde \theta_i + \gamma_3 \tilde \theta_{i-1} = \gamma_4 \tilde u_{i+1} + \gamma_5 \tilde u_{i-1}  + \gamma_6 \tilde v_i.\label{eq:tdisc}
 \end{eqnarray}
The coefficients $\alpha_k$, $\beta_k$ and $\gamma_k$ depend on the coefficients $A_k$, $B_k$ and $C_k$ and are given in appendix \ref{app:fdbrecoeff}. These finite-difference approximations are then encoded in a 3$\times$3 block matrix $A$ where each block is of size $(N+2)^2$. The first, second and third block rows contain the discretisations of the $x$-momentum, $y$-momentum and temperature equations respectively. To find the solution $\boldsymbol{\tilde u} = (\tilde u(\xi_i), \tilde v(\xi_i), \tilde \theta(\xi_i))^{T}$ for $0 \leq i \leq N+1$ we solve $A \boldsymbol{\tilde u} = \boldsymbol b$, where $b$ contains the values of the solution and its derivatives at the boundaries. 

 \begin{figure}
	\centering
	(a)\includegraphics[scale=0.3]{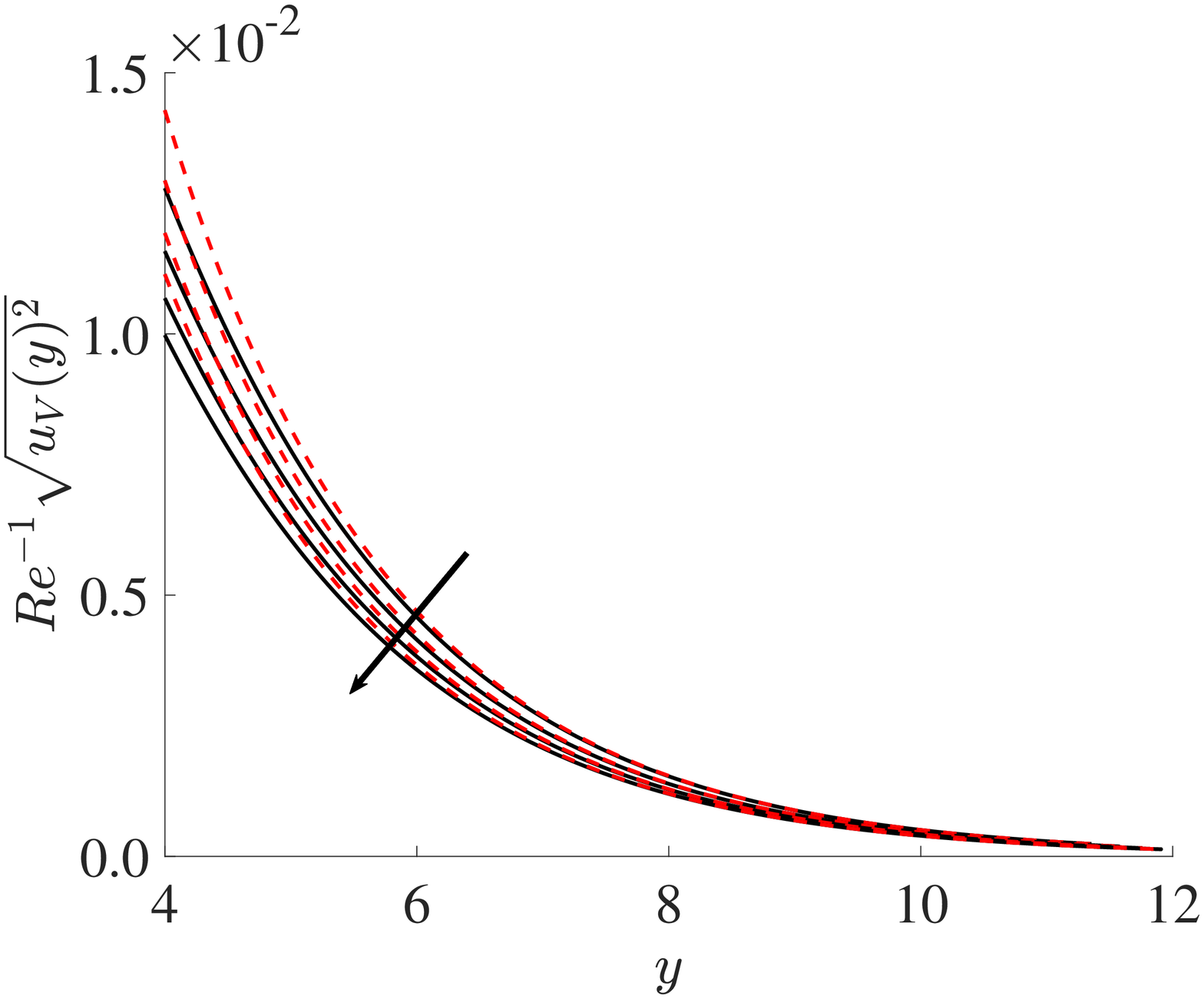}\label{fig:1a}	(b)\includegraphics[scale=.3]{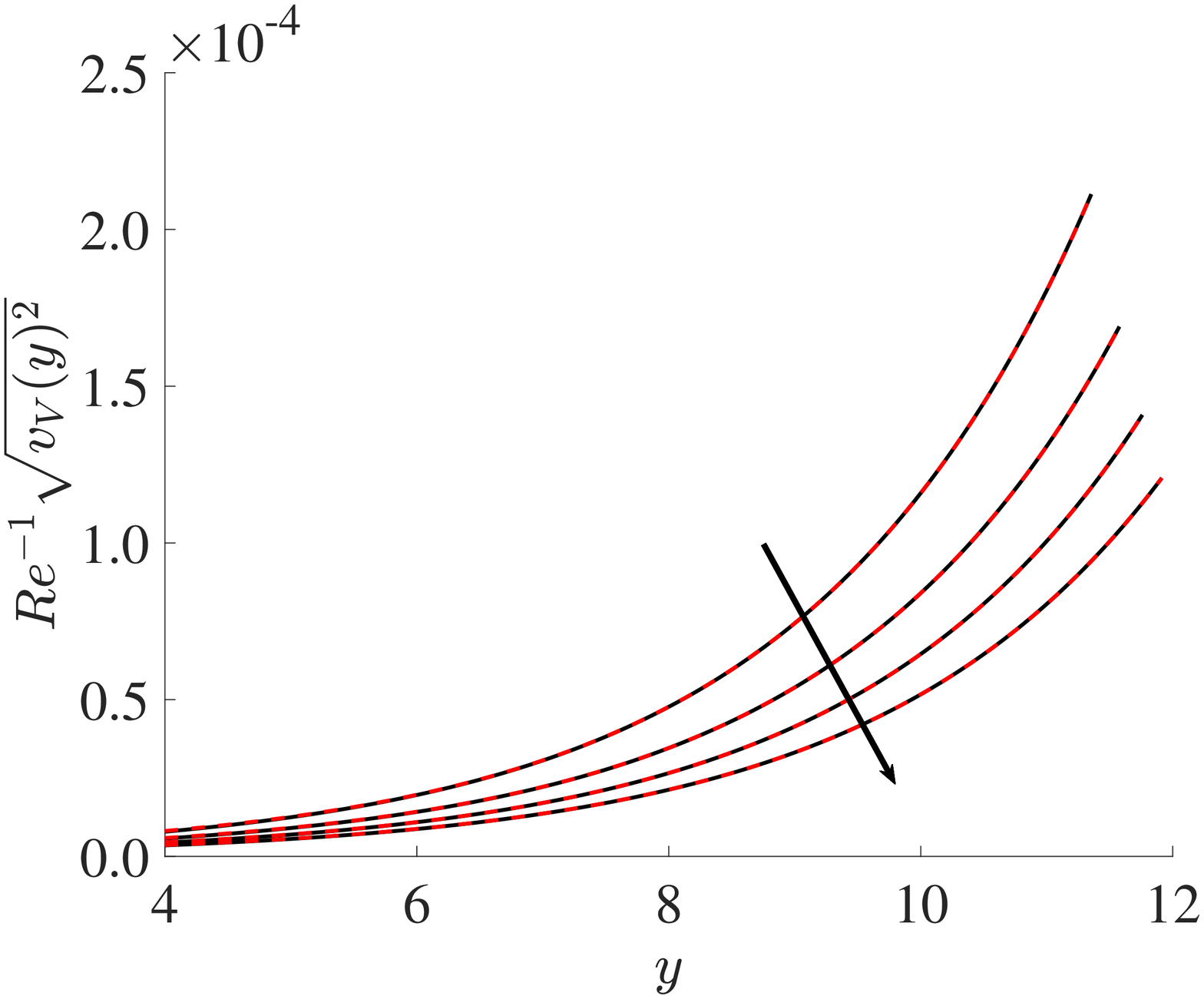}\label{fig:1b}
	(c)\includegraphics[scale=0.3]{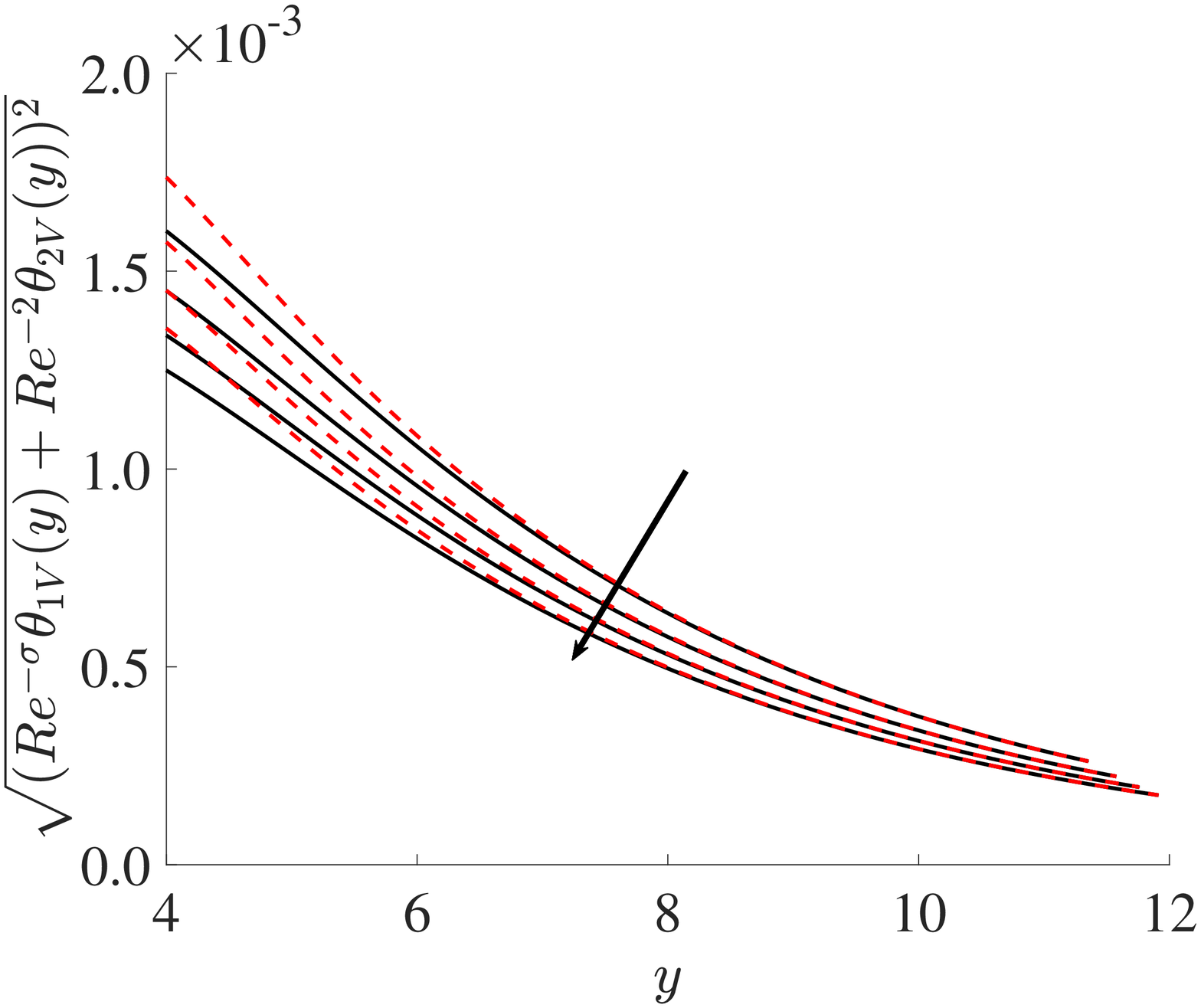}\label{fig:1c}
	\caption{The $z$-dependent part of $(a)$ the streak disturbance $u_v(Y)$, $(b)$ the normal velocity disturbance $v_v(Y)$ and $(c)$ the temperature disturbance $\theta_v(y)$, for Prandtl number $\sigma = 0.71$ (which is appropriate for air) and Reynolds number $Re = 80,000$, and free-stream Mach number $M_\infty = 0.8$. For comparison of the relative sizes we plot the Reynolds-number-scaled amplitude of each disturbance $\Rey \sqrt{u_v^2}$, $\Rey \sqrt{v_v^2}$ and $\sqrt{(\Rey^{-\sigma}\theta_{1v}(y)+ \Rey^{-2}\theta_{2v}(y) )^2}$ respectively. Red dashes denote the asymptotic solution from (\ref{eq:velsol1}), (\ref{eq:velsol2}) and (\ref{eq:tsol}). Black lines denote the numerical solution of the boundary-region equations (\ref{eq:breXmom})--(\ref{eq:breTemp}).}
	\label{fig:1}
\end{figure}
\section{Results}\label{sec:results}
 We solve the matrix system $A \boldsymbol{\tilde u} = \boldsymbol b$ on a grid containing $N=2000$ points. To compute the boundary conditions (\ref{eq:velsol1}), (\ref{eq:velsol2}) and (\ref{eq:tsol}), we require the value of $K_1 = K_1(\alpha, \beta)$ which is determined as part of the numerical solution of the production layer nonlinear eigenvalue problem (\ref{eq:PLmomen})--(\ref{eq:PLcontin}) for the wave speed $c_1$. For wavenumber values $(\alpha, \beta) = (0.2, 0.4)$, which by table \ref{tab:betasig} is in the regime where both the velocity and thermal streaks are expected to grow, \cite{deguchi_hall_2014a} find $K_1 = 16.9$; we use these parameter values in our computations.\\
 
\begin{figure}
	\centering
	(a)\includegraphics[scale=0.25]{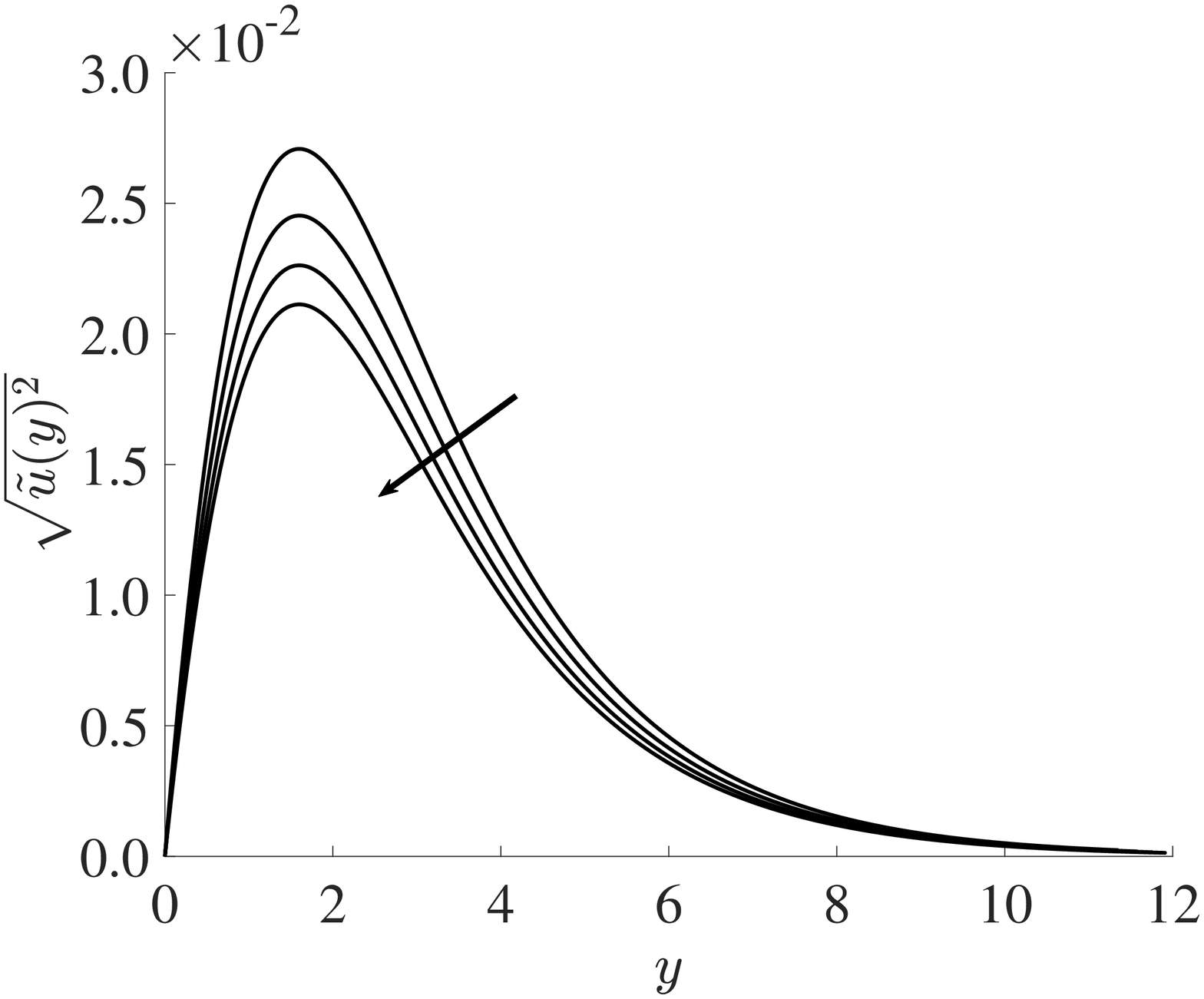}\label{fig:2a}
	(b)\includegraphics[scale=.25]{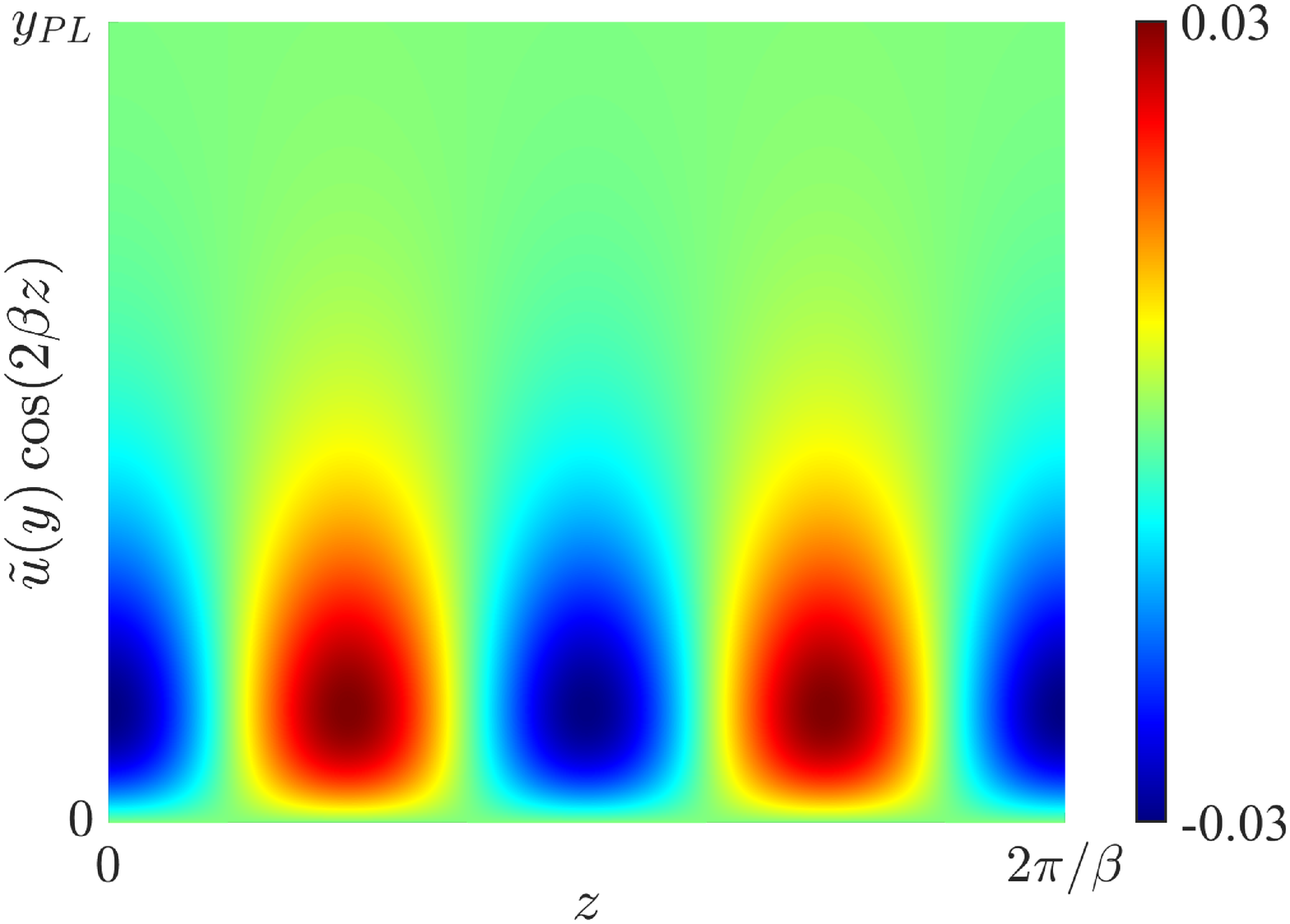}\label{fig:2b}
	(c)\includegraphics[scale=0.25]{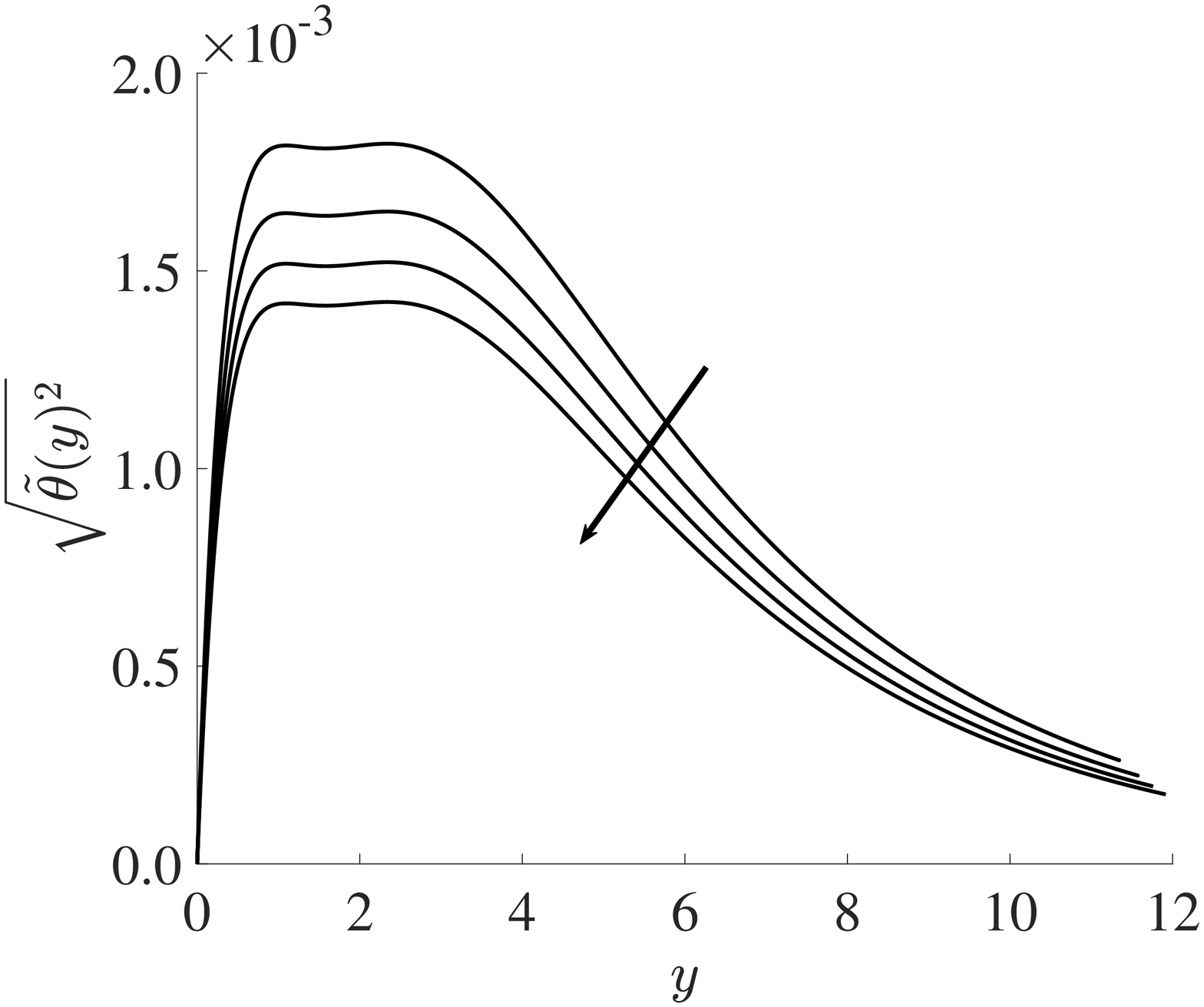}\label{fig:2c}	(d)\includegraphics[scale=.25]{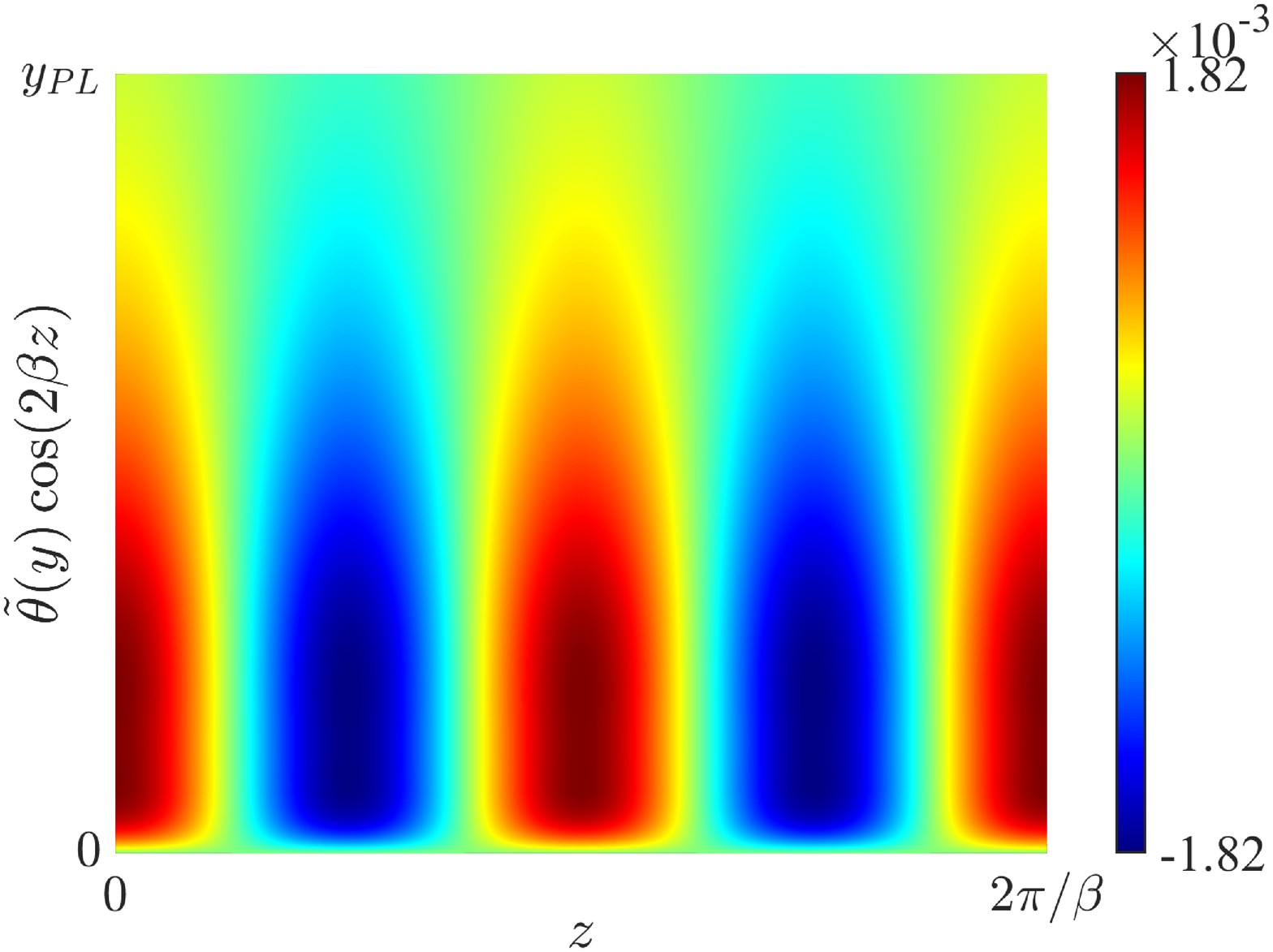}\label{fig:2d}
	
	\caption{The numerical solution to the boundary-region equations (\ref{eq:breXmom})--(\ref{eq:breTemp}) showing the streamwise velocity and temperature disturbances to the basic flow, $\tilde u$ and $\tilde \theta$. (a), (c): the amplitude $\sqrt{*^2}$ of the disturbances, for Prandtl number $\sigma = 0.71$, free-stream Mach number $M_\infty = 0.8$ and Reynolds numbers $\Rey = 80,000$, $100,000$, $120,000$ and $140,000$; the arrows indicate the direction of increasing Reynolds number. (b), (d): the velocity and thermal streaks $(b)$ $\tilde u(y) \cos(2 \beta z)$ and $(d)$ $\tilde \theta(y)\cos(2\beta z)$, shown over one period.}
	\label{fig:2}
\end{figure}
The amplitudes of the $z$-dependent parts of the streak, roll and temperature disturbances exiting the production layer are shown in figure \ref{fig:1}, together with the numerical solution of the boundary-region equations. The asymptotic solution describing the normal velocity $v_v$ is valid all the way to the wall, whereas the asymptotic solutions for the exponentially growing velocity and thermal streaks start to visibly break down at around $y=7$. As the wall is approached, the solution of the boundary-region equations describes the flow induced by the disturbances from the production layer. 

The solution of the numerical boundary-region equations across the main part of the boundary-region to the wall is shown in figure \ref{fig:2}. As in the incompressible case, the velocity streak grows throughout the boundary region before taking its maximum in the near-wall boundary layer. In the compressible problem, the nonlinear interaction also produces a thermal streak which similarly grows throughout the boundary layer; the rate growth of the thermal streak is higher than that of the velocity streak. The effect of the thermal streak is also felt more uniformly across the flow compared to the velocity streak.

\begin{figure}
	\centering
	(a)\includegraphics[scale=0.27]{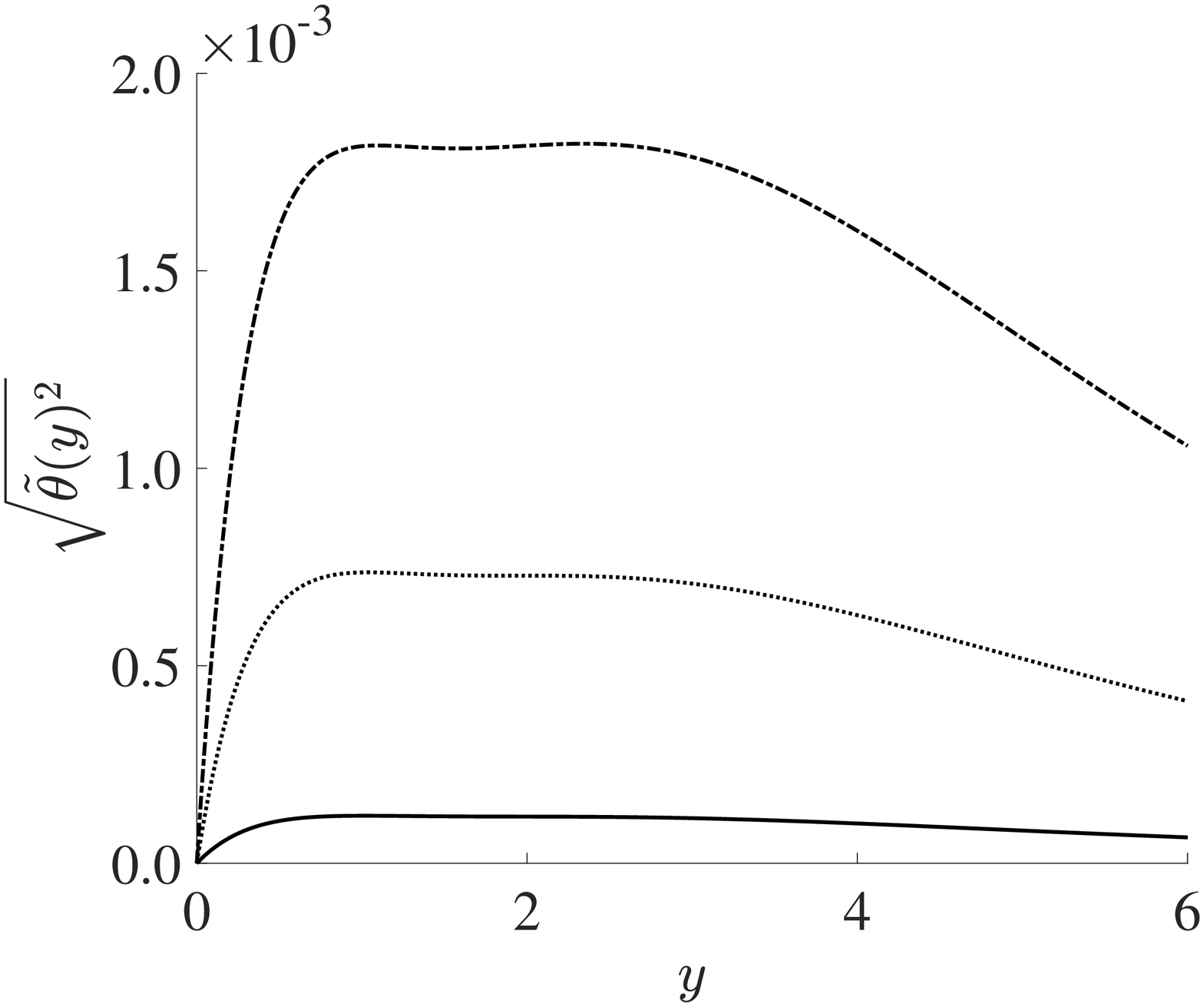} (b)\includegraphics[scale=0.27]{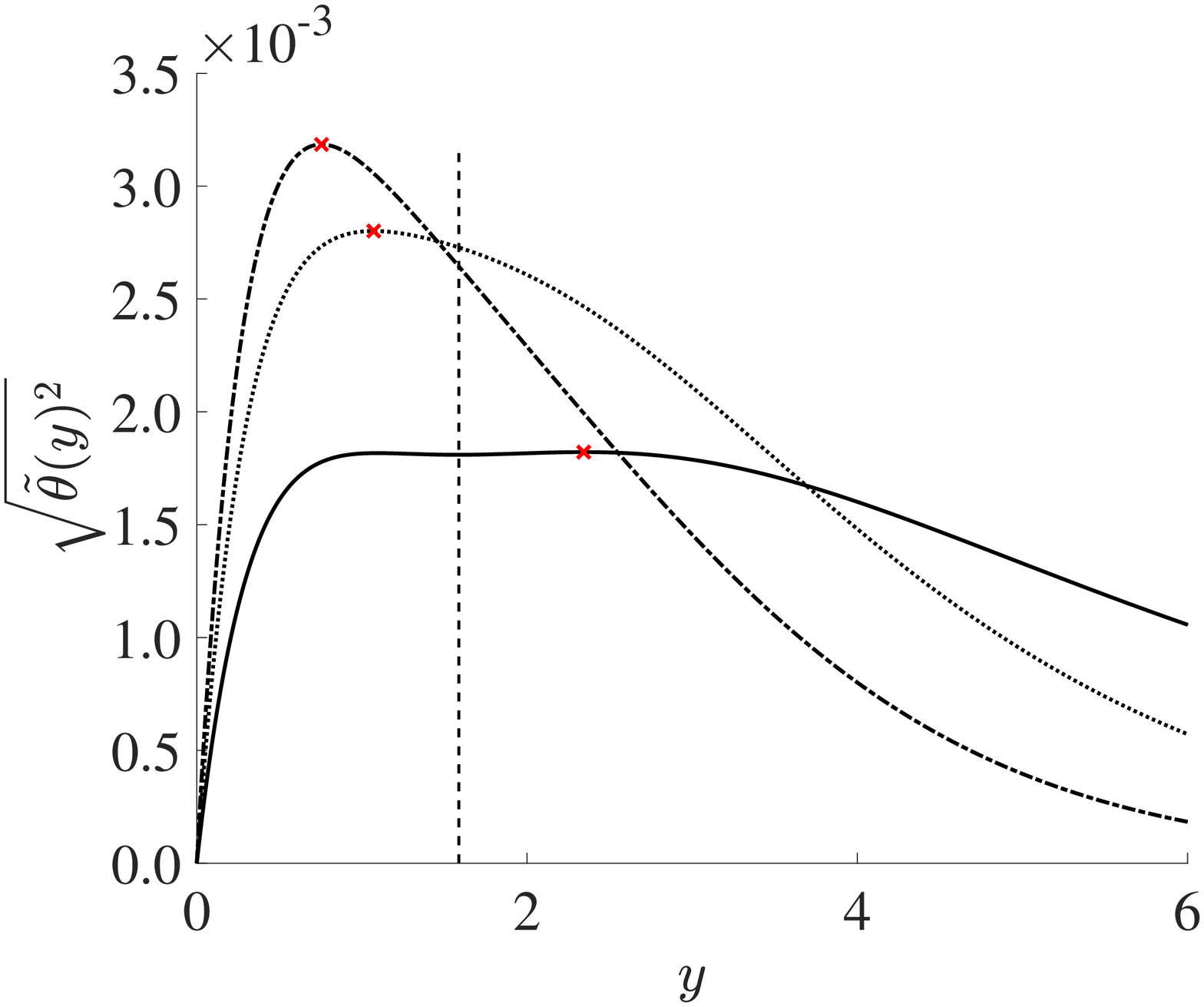}
	
	\caption{(a): the amplitude of the thermal streak $\tilde \theta$ for Reynolds number $\Rey = 80,000$, Prandtl number $\sigma = 0.71$ and free-stream Mach numbers $M_\infty = 0.2$ (solid curve), $M_\infty =0.5$ (dotted curve) and  $M_\infty =0.8$ (dot-dashed curve). (b): the amplitude of the thermal streak $\tilde \theta$ for Reynolds number $\Rey = 80,000$, free-stream Mach number $M_\infty = 0.7$ and for Prandtl numbers $\sigma = 0.71$, $\gamma = 1.4$ (solid curve), $\sigma = 1$, $\gamma=1.4$ (dotted curve) and $\sigma = 1.3$, $\gamma=1.3$ (dot-dashed curve). The red crosses show the location of the maximum amplitude. The $y$-location of the maximum value of the streak disturbance $\tilde u$ is indicated by the dashed line.}
		\label{fig:3}
\end{figure}

The variation of the amplitude of the thermal streak for varying Mach number and Prandtl number is shown in figure \ref{fig:3}, for a Reynolds number of $\Rey = 80,000$. The amplitude of the thermal streak is enhanced as the free-stream Mach number is increased, which is consistent with the idea of compressibility effects becoming more important as the free-stream Mach number is increased \citep{morkovin1962effects}. Additionally, the variation of the free-stream Mach number has a much larger effect on the amplitude of the thermal streaks than the variation of the Reynolds number. Meanwhile the effect of varying the Prandtl number is shown in figure \ref{fig:3}(b), where we show the amplitude of the thermal streak for Prandtl numbers $\sigma = 0.71$, $1$ and $1.3$, together with the $y$-location of the maximum value of the streak disturbance. For $\sigma = 0.71$ and $\sigma = 1$ we use a heat capacity ratio $\gamma = 1.4$; for $\sigma = 1.3$ which is approximately the Prandtl number for gaseous ammonia we use a heat capacity ratio of $\gamma = 1.3$. The maximum value of the temperature disturbance occurs closer to the wall as the Prandtl number $\sigma$ increases, and for $\sigma \geq 1$ occurs closer to the wall than that of the streak disturbance. Thus if thermal diffusivity dominates, the thermal streaks dissipate faster than the velocity streaks, and vice-versa if momentum diffusivity dominates. 

\section{Discussion}\label{sec:discussion}
Our results show the existence of free-stream coherent structures in the compressible asymptotic suction boundary layer at $O(1)$ Mach number. The solutions take the form of a roll--wave--streak interaction at the edge of the boundary layer, in a production layer whose location is dependent on both the Prandtl number and the Mach number. The interaction produces both a streaky disturbance and a temperature disturbance. These grow exponentially out of the production layer, with the rate of growth being controlled by the spanwise wavenumber and, for the temperature disturbance, the Prandtl number. Above the layer, the disturbances decay rapidly to zero. For the compressible case considered here, the main difference from the incompressible case is the development of a spanwise varying temperature field beneath the production layer. The size of the induced temperature field depends on the Prandtl number. The latter temperature field and the induced streak may well lead to secondary instabilities in the main part of the boundary layer. In the incompressible case we know from \citet*{dempsey2017excitation} that the streak generated by the free-stream coherent structure acts as a receptivity mechanism in curved flows, so for curved compressible flows, such as those over turbine blades, we anticipate that the structures described here might trigger transition through the G\"ortler vortex mechanism.

Our results show that the fundamental mechanism described by \citet{deguchi_hall_2014a} for incompressible flows is also operational in compressible flows. In particular, this suggests the the mechanism will occur in compressible jets and therefore might have important consequences for sound production in compressible jet flows. Extension of the work of \citet{deguchi_hall_2015a} on swept wing flows is also possible.  

Our analysis has assumed that the viscosity can be described by the Chapman--Rubesin law \citep{chapman1949temperature} and that the gas in question is an ideal gas. Extension of the work to account for a more realistic viscosity model, for example Sutherland's law \citep{sutherland1893lii}, is straightforward but we believe that for the Mach numbers considered here that is not necessary. At hypersonic speeds beyond the regime covered here both real gas effects and more realistic viscosity models must be used and an intriguing question is the relationship between the production layer problem and the temperature adjustment layer for the basic state at hypersonic speeds. Certainly, we know from for example \citet*{cowley1190instability}, \citet{blackaby_cowley_hall_1993} and \cite{fu1993gortler} that real gas effects, realistic viscosity models and indeed shocks present in the flow can significantly alter streamwise vortex or travelling wave instabilities, so it is to be expected that the free-stream coherent structure mechanism at hypersonic speeds will be significantly different from that in  the incompressible case. 

It is not yet known whether the class of exact coherent structures described by \cite{hall_sherwin_2010} can be extended to compressible flows. However the fundamental asymptotic analysis  supporting the structure is the vortex-wave interaction theory of \cite{hall_smith_1991} which in fact was given in the context of compressible flows so it would appear likely that it is operational in compressible flows. Moreover, the inviscid stability equation for many boundary layer compressible flows has unstable solutions when the incompressible counterpart has none so it may well be that vortex-wave interactions in compressible flows may have a richer structure than their incompressible counterparts. Taken together with the extension of the free-stream coherent structure mechanism to compressible flows, it suggests that compressible flows might well have a significantly family of possible exact coherent states.

Declaration of Interests. The authors report no conflict of interest.

\appendix
\section{Coefficients of the finite-difference approximation to the boundary-region equations}\label{app:fdbrecoeff}
The coefficients of the finite-difference approximation to the $x$-momentum equation (\ref{eq:breXmom}) are given by
\begin{subeqnarray}
	&\alpha_1 =  A_2/\Delta \xi + A_3 /\Delta \xi^2, \quad \alpha_2 = A_1 - 2A_3/\Delta \xi^2, \quad \alpha_3 = -A_2/\Delta \xi + A_3 /\Delta \xi^2, \\
	&\alpha_4 = A4, \quad \alpha_5 = A_6/2\Delta \xi, \quad \alpha_6 = A_5, \quad \alpha_7 = -A_6/2\Delta \xi, \\
\end{subeqnarray}
The coefficients of the finite-difference approximation to the $y$-momentum equation (\ref{eq:breYmom}) are given by
\begin{subeqnarray}
	& \beta_1 = B4/2\Delta \xi^3 + B5/\Delta \xi^4, \\
	& \beta_2 = B2/2\Delta \xi + B3/\Delta \xi^2 -2B4/2\Delta \xi^3 -4B5/\Delta \xi^4, \\
	& \beta_3 = B1 - 2B3/\Delta \xi^2 + 6B5/\Delta \xi^4, \\
	& \beta_4 = -B2/2\Delta \xi + B3/\Delta \xi^2 + 2B4/2\Delta \xi^3 -4B5/\Delta \xi^4,\\
	&\beta_5 = -B4/2\Delta \xi^3 + B5/\Delta \xi^4, \quad
	 \beta_6 = B9/2\Delta \xi^3 + B10/\Delta \xi^4, \\
	& \beta_7 = B7/2\Delta \xi + B8/\Delta \xi^2 -2B9/2\Delta \xi^3 -4B10/\Delta \xi^4, \\
	& \beta_8 = B6 - 2B8/\Delta \xi^2 + 6B10/\Delta \xi^4, \\
	& \beta_9 = -B7/2\Delta \xi + B8/\Delta \xi^2 + 2B9/2\Delta \xi^3 -4B10/\Delta \xi^4,\\
	&\beta_10 = -B9/2\Delta \xi^3 + B10/\Delta \xi^4,\\
\end{subeqnarray}
The coefficients of the finite-difference approximation to the temperature equation (\ref{eq:breTemp}) are given by
\begin{subeqnarray}
	&\gamma_1 = C1/2\Delta \xi, \quad \gamma_2 = -C1/2\Delta \xi, \quad \gamma_3 = C2, \quad
	\gamma_4 = C4/2\Delta \xi + C5/\Delta \xi^2, \\
	& \gamma_5 = C3 -2C5/\Delta \xi^2, \quad  \gamma_6 = -C4/2\Delta \xi +C5/\Delta \xi^2.
\end{subeqnarray}
The coefficients A1-A6, B1-B10 and C1-C5 are available from the authors on request. 
\bibliographystyle{jfm}
\bibliography{transbib}

\end{document}